\documentclass[aps,pra,amssymb,superscriptaddress,twocolumn]{revtex4-1}
\usepackage[dvipdfmx]{graphicx}
\usepackage{bm}
\usepackage{amsmath}
\usepackage{ascmac}
\usepackage{color}
\usepackage{setspace}
\usepackage{amssymb}
\usepackage{latexsym}
\usepackage{mathtools}
\usepackage{lipsum}
\usepackage[colorlinks,linkcolor=blue,citecolor=blue]{hyperref}

\begin{document}

\title{Scale-invariant relaxation dynamics in two-component Bose-Einstein condensates with large particle-number imbalance} 

\author{Kazuya Fujimoto}
\affiliation{Institute for Advanced Research, Nagoya University, Nagoya 464-8601, Japan}
\affiliation{Department of Applied Physics, Nagoya University, Nagoya 464-8603, Japan}

\author{Kazunori Haneda}
\affiliation{Department of Applied Physics, Nagoya University, Nagoya 464-8603, Japan}

\author{Kazue Kudo}
\affiliation{Department of Computer Science, Ochanomizu University, Tokyo 112-8610, Japan}

\author{Yuki Kawaguchi}
\affiliation{Department of Applied Physics, Nagoya University, Nagoya 464-8603, Japan}

\begin{abstract}
We theoretically study the scale-invariant relaxation dynamics in segregating two-component Bose-Einstein condensates with large particle-number imbalance, and uncover that random walk of droplet for the minor component plays a fundamental role in the relaxation process. Our numerical simulations based on the binary Gross-Pitaevskii model reveal the emergence of the dynamical scaling during the relaxation, which is a hallmark of scale-invariant dynamics, in a correlation function for the minor condensate. Tracking exponents characterizing the dynamical scaling in time, we find out a crossover phenomenon that features the change in power exponents of the correlation length. To understand the fundamental mechanism inherent in the scale-invariant relaxation dynamics, we construct a random walk model for droplets. Employing the model, we analytically derive the $1/3$ and $1/2$ power laws and predict the crossover of the scaling. These exponents are in reasonable agreement with the values obtained in the numerical calculations. We also discuss a possible experimental setup for observing the scale-invariant dynamics.
\end{abstract}

\date{\today}

\maketitle

\section{Introduction}
Ultracold atomic gases have been one of the ideal playgrounds for studying nonequilibrium phenomena in quantum systems because of the high tunability of physical parameters and  long enough time scales for observing the dynamics in experiments \cite{coldgas1,coldgas2,coldgas3}. In fact, a variety of fundamental issues such as thermalization \cite{thermalization1,thermalization2,thermalization3}, phase transition \cite{transition1,transition2,transition3,KZM1,KZM2,KZM3,KZM4,KZM5}, and turbulence \cite{turbulence1,turbulence2,turbulence3,turbulence4,turbulence5} have been investigated. Among them, universal aspects of nonequilibrium phenomena have been attracting a lot of attention. A typical example is scale-invariant relaxation dynamics, features of which are captured by a correlation function $C(r,t)$ that exhibits dynamical scaling $C(r,t) = s^{-\gamma} C(r s^{\beta}, t s)$ with a constant $s$ \cite{cardy,hohenberg,onuki2002,bray2002,tauber2014}. The exponents $\beta$ and $\gamma$ classify universality of relaxation dynamics. The recent works in ultracold quantum gases reveal the existence of several universality classes in terms of coarsening dynamics \cite{Sachdev1,Coarsening_de,Coarsening_kudo1,Coarsening_kudo2,Coarsening_kudo3,Coarsening_Hofmann,Coarsening_blakie1,Coarsening_blakie2,Coarsening_blakie3,Coarsening_blakie4,Coarsening_blakie5,Coarsening_blakie6,Coarsening_take1,Coarsening_take2,Coarsening_take3,Coarsening_take4,Coarsening_de,Coarsening_phuc,Coarsening_fuji,Coarsening_Proukakis1,Coarsening_Proukakis2,arXiv1,arXiv2} and non-thermal fixed points (NTFPs)\cite{Nowak1,Nowak2,Schole1,Karl1,Kral2,Karl3,Schmied1,NTFP_fuji, NTFP_exp1,NTFP_exp2}. In the context of coarsening dynamics, segregating binary Bose-Einstein condensates (BECs) are shown to have $\beta=2/3$ and belong to the universality class for classical binary liquids in a inertial regime \cite{bray2002} when the particle numbers in the two condensates are equal \cite{Coarsening_kudo2,Coarsening_Hofmann,Coarsening_blakie1}. 

This kind of universal relaxation with particle-number balance has been originally studied in classical binary liquid at a critical quench for a long time \cite{onuki2002,bray2002}. However, many systems with particle-number imbalance have given another essential testbed for comprehending scale-invariant relaxation dynamics. In fact, Ostwald ripening, which is an important phenomenon in such a system, has been actively studied. In this phenomenon, many small droplets of the minor component are nucleated and subsequently grow or shrink in time. One of the established theories is originally developed by Lifshitz and Slyozov \cite{Ostwald0} and Wagner \cite{Ostwald1}, who uncovered that diffusion processes (evaporation and concentration) are of importance in the dynamics and that the averaged droplet size is governed by a $\beta =1/3$ power law. The original idea for the theory has been improved and extended to other systems by many researchers \cite{Ostwald2,Ostwald3,Ostwald4,Ostwald5,Ostwald6}, and thus Ostwald ripening has played a fundamental role in developing a basis for the universal dynamics.

Against its backdrop, we bring up an interesting question; ``Does a kind of Ostwald ripening emerge even in quantum gases?" Usually, ultracold atomic gases are well separated from the environment, and are regarded as isolated quantum systems. Thus, the total energy is conserved in contrast to the corresponding classical systems, and the diffusion process mentioned above is generally prohibited.
This implies that droplets cannot shrink nor grow through the Ostwald ripening mechanism. On the other hand, coarsening dynamics actually proceeds even in the absence of diffusive processes in a binary BEC with particle-number balance. We expect that similar coarsening dynamics also occurs in a particle-number imbalanced system.  It is then natural to ask what kind of universal phase-ordering kinetics emerges.

In this paper, we theoretically find that the particle-number imbalance greatly affects the universal aspects of the relaxation dynamics in two-dimensional (2D) segregating binary BECs, and demonstrates that the relaxation mechanism is essentially different from Ostwald ripening in classical systems. Our theoretical model is a two-component Gross-Pitaevskii (GP) equation, and we firstly perform the numerical simulations starting from initial states generated by the truncated Wigner method \cite{TWA1,TWA2,TWA3}. Then, we reveal that many droplets of the minor condensate are formed and that the correlation function of the wavefunction for the minor component shows dynamical scaling. Furthermore, we numerically find a crossover of the power law: the exponent $\beta$ is close to $1/3$ in the early stage of the dynamics and approaches to $1/2$ in the late time. To unveil a mechanism behind the scale-invariant dynamics, we employ an idea that random collisions between droplets promote the relaxation dynamics, and then analytically derive $1/3$ and $1/2$ power laws for the correlation length, which well explain the numerical results. Therefore, we conclude that the droplet random walk is the origin of the power laws, and argue that the mechanism is quite different from Ostwald ripening in classical systems in which the diffusion processes are crucial. Here, it is noteworthy that there are a few theoretical works related to a percolation problem in binary BECs with the imbalance \cite{Coarsening_take1,Coarsening_take2,Coarsening_blakie6}, but the several things mentioned above is not yet to be considered.

The rest of paper is organized as follows. In Sec.~\ref{formulation}, we describe our theoretical tools such as the binary GP model, a tuning protocol for the particle-number imbalance, and the dynamical scaling in a correlation function. Section~\ref{Num} shows our numerical results, and we find that the particle-number imbalance gives a significant impact on the scale-invariant relaxation. To understand the universal aspects of the relaxation, in Sec.~\ref{ana_derivation}, we analytically obtain two power exponents for the growth laws of the correlation length, and unveil the fundamental relaxation mechanism. In Sec.~\ref{discussion}, we explain the relation between the universal relaxation dynamics found here and previous literature, and discuss an experimental setup for observing our theoretical prediction. Finally, we summarize our results in Sec.~\ref{Conclusion}.

\section{Theoretical formulation \label{formulation}}
\subsection{Model}
We consider a two-component BEC close to zero temperature in a 2D uniform system. The system is well described by two macroscopic wavefunctions $\psi_{m}(\bm{r},t)~(m=1,2)$ obeying the following binary Gross-Pitaevskii (GP) equation \cite{pethick,stringari}: 
\begin{widetext}
\begin{eqnarray}
i\hbar \frac{\partial}{\partial t}\psi_{1} (\bm{r},t) &=& -\frac{\hbar^2}{2M} \bm{\nabla}^2 \psi_{1}(\bm{r},t) + g_0 |\psi_{1}(\bm{r},t)|^2 \psi_{1}(\bm{r},t) + g_1 |\psi_{2}(\bm{r},t)|^2 \psi_{1}(\bm{r},t) + \frac{ \hbar \Omega(t)}{2} \psi_{2}(\bm{r},t), \label{GP1} \\
i\hbar \frac{\partial}{\partial t}\psi_{2}(\bm{r},t) &=& -\frac{\hbar^2}{2M} \bm{\nabla}^2 \psi_{2}(\bm{r},t) + g_0 |\psi_{2}(\bm{r},t)|^2 \psi_{2}(\bm{r},t) + g_1 |\psi_{1}(\bm{r},t)|^2 \psi_{2}(\bm{r},t) + \frac{ \hbar \Omega(t)}{2} \psi_{1}(\bm{r},t)  \label{GP2}
\end{eqnarray}
\end{widetext}
with a particle mass $M$, intra- and inter-species interaction couplings $g_0>0$ and $g_1>0$, and a Rabi coupling $\Omega (t)$. The ground state with $\Omega(t)=0$ is immiscible (miscible) for $g_1>g_0$ ($g_1<g_0$), in which the two condensates are separated into two spatially distinct regions (are overlapped and spread in the whole region). In this work, we assume an immiscible binary system, i.e., $g_1>g_0$. As explained in Sec.~\ref{protocol}, the Rabi coupling $\Omega(t)$ is used to adjust a particle-number imbalance parameter $R=N_2/N_1$. Here, the particle number $N_m$ for the component $m$ is defined by
\begin{eqnarray}
N_{m} = \int_{\mathcal{S}} |\psi_{m}(\bm{r},t)|^2 d\bm{r} \label{GP3}
\end{eqnarray}
with the whole system space $\mathcal{S}$.

\begin{figure*}[t]
\begin{center}
\includegraphics[keepaspectratio, width=18cm,clip]{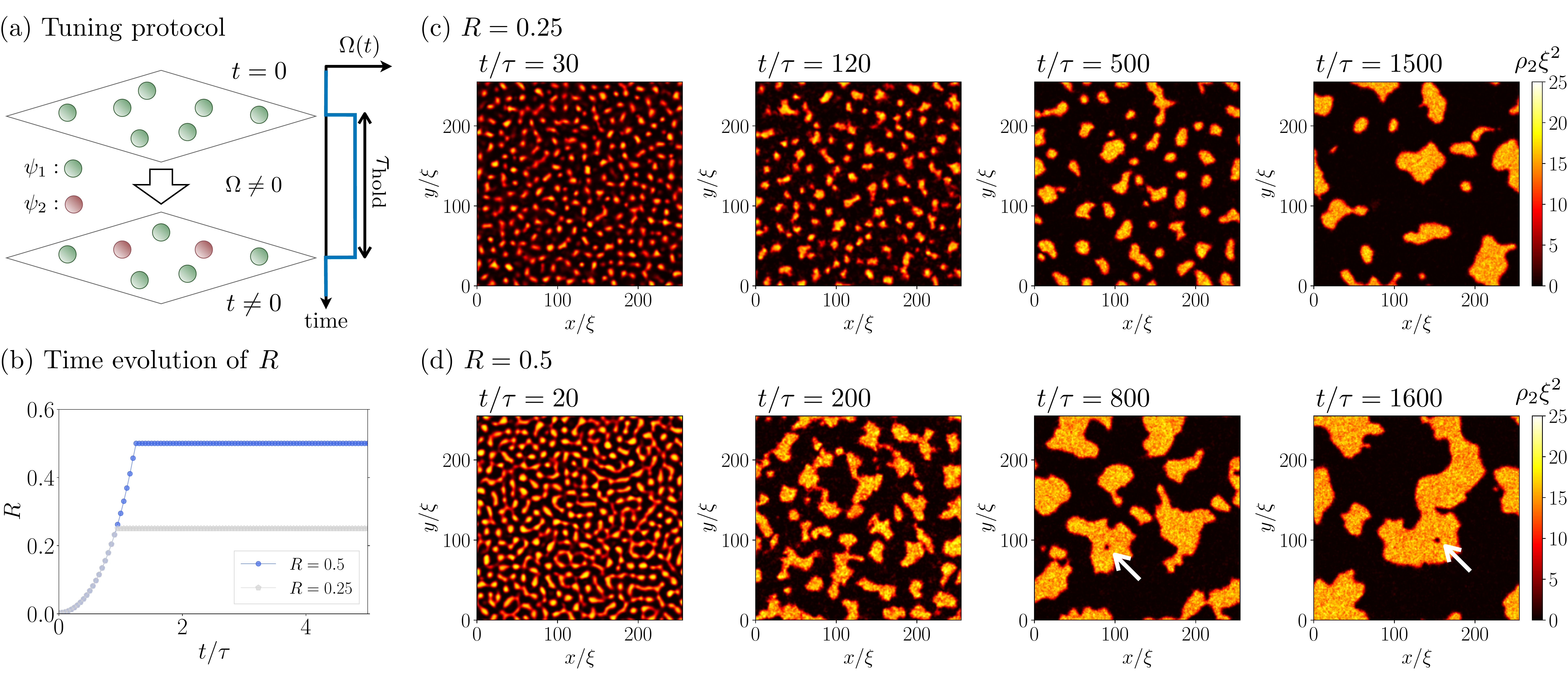}
\caption{(Color online) (a) Schematic of the tuning protocol to adjust the particle-number imbalance parameter $R$. In the initial state, only component $1$ is condensed and there are tiny fluctuations on it due to the Bogoliubov vacuum (see the detail of Sec.~\ref{initial_state}). At $t=0$, we suddenly switch on the Rabi coupling $\Omega (t)$, and a particle number of the component $2$ then increases in time. Just after $R(t)$ reaches a specified value, we suddenly turn off the coupling. The time evolution of $\Omega(t)$ in the whole process is depicted in the right side, which corresponds to Eq.~\eqref{GP4}.  (b) Time evolution of $R(t)$ for two specified values in numerical calculations. The imbalance parameter rapidly grows during the early stage of the dynamics, and becomes constant just after turning off the Rabi coupling. (c,d) Snapshots of the density distributions $\rho_{2}(\bm{r},t)=|\psi_2(\bm{r},t)|^2$ for $R=0.25$ (c) and $0.5$ (d). Figures (c) and (d) exhibit that the circular droplet structures can survive for a longer time for smaller $R$. In (d), composite vortices are denoted by solid arrows (see text).
\label{fig1} }
\end{center}
\end{figure*}

\subsection{How to prepare initial states \label{initial_state}}
Our system is a 2D square box $L \times L$ with the periodic boundary condition. An initial wavefunction of Eqs.~\eqref{GP1} and \eqref{GP2} is prepared by employing the truncated Wigner method with a Bogoliubov vacuum \cite{TWA1,TWA2,TWA3}. We consider a single condensed state of the component $1$ as the vacuum. Then, we generate the initial state by using the following expressions with random variables $\alpha_{j,\bm{k}}~(j=1,2)$:
\begin{eqnarray}
\psi_1(\bm{r},0) &=& \sqrt{\rho_{\rm tot}} \nonumber \\
&+& \frac{1}{\sqrt{S}} \sum_{\bm{k} \in \{ \bm{k} | E_{\bm{k}} \leq \mu, \bm{k} \neq 0 \} } \bigl( u_{\bm{k}}\alpha_{1,\bm{k}} + v_{\bm{k}}\alpha_{1,-\bm{k}}^{*}  \bigl) e^{i\bm{k}\cdot\bm{r}},  \nonumber \\
\label{GP5_1}
\end{eqnarray}
\begin{eqnarray}
\psi_2(\bm{r},0) = \frac{1}{\sqrt{S}} \sum_{\bm{k} \in \{ \bm{k} | \epsilon_{\bm{k}} \leq \mu, \bm{k} \neq 0 \}} \alpha_{2,\bm{k}} e^{i\bm{k}\cdot\bm{r}}, \label{GP5}
\end{eqnarray}
\begin{eqnarray}
u_{\bm{k}} = \sqrt{ \frac{\epsilon_{\bm{k}} + \mu}{2E_{\bm{k}}}+ \frac{1}{2}},  \label{GP6}
\end{eqnarray}
\begin{eqnarray}
v_{\bm{k}} = - \sqrt{ \frac{\epsilon_{\bm{k}} + \mu}{2E_{\bm{k}}}- \frac{1}{2}},  \label{GP7}
\end{eqnarray}
\begin{eqnarray}
E_{\bm{k}} = \sqrt{ \epsilon_{\bm{k}}(\epsilon_{\bm{k}} + 2\mu )}, \label{GP8}
\end{eqnarray}
where $\rho_{\rm tot}=N_0/S$ is the number density of the condensed particles with $N_0$ being the number of condensed particles and $S=L^2$ the system area, $\mu=g_0\rho_{\rm tot}$ is the chemical potential, and $\epsilon_{\bm{k}} = \hbar^2 k^2 /2M$ is the single-particle kinetic energy.
The complex random numbers $\alpha_{j,\bm{k}}$ are sampled by the Wigner function $W( \{ \alpha_{j,\bm{k}} \})$ corresponding to the Bogoliubov vacuum: 
\begin{eqnarray}
W( \{ \alpha_{j,\bm{k}} \}) = A  \prod_{\bm{k}} {\rm exp} \Bigl( -2 | \alpha_{j,\bm{k}} |^2  \Bigl)  \label{GP9}
\end{eqnarray}
with the normalization factor $A$.

\subsection{Protocol for tuning the particle-number imbalance \label{protocol}}
We study how the particle-number imbalance affects the dynamical scaling behavior of a segregating binary mixture in the immiscible parameter regime. For this purpose, we use a protocol to tune $R$ as shown in Fig.~\ref{fig1}(a), where the Rabi coupling is used to adjust the imbalance parameter $R$. Our work assumes the coupling $\Omega(t)$ is given by 
\begin{eqnarray}
 \Omega(t)= \frac{1}{\tau} \theta(\tau_{\rm hold}-t) \theta(t)  \label{GP4}
\end{eqnarray}
with the holding time $\tau_{\rm hold}$, the characteristic time $\tau = \hbar/\mu$, and the step function $\theta(x)$. Under the tuning protocol, the total particle number $N_{\rm tot}=N_1 + N_2$ is conserved in time, and we can freely tune the parameter $R$ as shown in Fig.~\ref{fig1}(b), where we show numerical results of Eqs.~\eqref{GP1} and \eqref{GP2} on how the imbalance parameter $R$ changes in time. This method has already been implemented in several experiments for studying non-equilibrium phenomena in binary BECs \cite{binary_exp1,binary_exp2,binary_exp3}. In this work, considering only $R<1$ in which the component $2$ is minority, we study relaxation dynamics, namely phase separation kinetics after the protocol. 

\subsection{Correlation function and dynamical scaling \label{correlation_review}}
Scale-invariant relaxation dynamics is well known to be captured by several correlation functions \cite{cardy,hohenberg,onuki2002,bray2002,tauber2014}. 
To quantify what extent our system exhibits scale-invariant properties, we introduce a correlation function for $\psi_{2}({\bm r},t)$:
\begin{eqnarray}
C(r,t) = \biggl{\langle}  \frac{1}{S}  \int_{\mathcal{S}}  \psi_2^*({\bm r} + {\bm x},t) \psi_2({\bm x},t) d\bm{x}  \biggl{\rangle}_{\Omega},  
\label{correlation1}
\end{eqnarray}
where $\langle \cdots\rangle_{\Omega}$ denotes an average over the angle of $\bm{r}$. 
The correlation function $C(r,t)$ takes only real values because of the angle average (see Appendix~\ref{imaginary} for the proof). We define a correlation length $L(t)$ as
\begin{eqnarray}
C(L(t),t) = 0.5C(0,t).  \label{correlation2}
\end{eqnarray}
The length $L(t)$ is a characteristic scale for the minor condensate and corresponds to a typical size of droplets generated after the tuning protocol as described in Sec.~\ref{Num}.

Here, we comment on a density correlation function for $\rho_2(\bm{r},t)$. 
This correlation function includes information for a domain distribution around a droplet because $\rho_2(\bm{r},t)$ always gives positive contribution. On the other hand, in Eq.~\eqref{correlation1}, the phase of the macroscopic wavefunction $\psi_2(\bm{r},t)$ on droplets are almost independent, and thus effects of the surrounding domains can be canceled due to the integral over space. Thus, Eq.~\eqref{correlation1} is suitable for investigating growth of the droplet size rather than using the density correlation function.

When the system has scale-invariant dynamics, the correlation function obeys the dynamical scaling:
\begin{eqnarray}
C(r,t) = s^{-\gamma} C(r s^{\beta}, t s)
\label{correlation4}
\end{eqnarray} 
with a variable $s$.
Substituting $s=t^{-1}$ in Eq.~\eqref{correlation4}, we obtain
\begin{eqnarray}
C(r,t) = t^{\gamma} G(r t^{-\beta})  
\label{correlation5}
\end{eqnarray} 
with a function $G(x) = C(x,1)$. This means that the correlation length $L(t)$ and the on-site correlation $C(0,t)$ have power laws:
\begin{eqnarray}
L (t)    & \propto &  t^{\beta},  \label{correlation6} \\
C(0,t) & \propto &  t^{\gamma}.
\end{eqnarray} 
In our system without the Rabi coupling, the particle number for each component is independently conserved in time, and thus the exponent $\gamma$ is zero. This fact is numerically confirmed in the next section.

This scaling law has been discussed in the context of critical dynamics close to second-order phase transitions in equilibrium states over several decades. However, several theoretical studies have recently found the emergence of such dynamical scalings in isolated quantum systems even without critical points, and the universality are investigated from the perspective of NTFPs \cite{berges2015,Nowak1,Nowak2,Schole1,Karl1,Kral2,Karl3,Schmied1,Berges1,Orioli1,Berges2,Aleksas1,NTFP_fuji}. So far, the two experiments \cite{NTFP_exp1,NTFP_exp2} report observations of the NTFP scenario in scalar and spinor Bose gases. 

\section{Numerical results \label{Num}}
We numerically study the relaxation dynamics with large particle-number imbalance by tuning several values of $R$ smaller than unity. This section describes scale-invariant aspects of the relaxation using the density distributions and the correlation functions. The parameters used in our numerics are given as follows: $\rho_{\rm tot} = 15/\xi^2$, $g_1 / g_0 = 1.5$ and $L=256 \xi$ with the coherence length $\xi=\hbar/\sqrt{2M \mu}$. Using the pseudo-spectral method with the Fourth-order Runge-Kutta time evolution, we numerically solve Eqs.~\eqref{GP1} and \eqref{GP2}. The numerical resolutions for space and time are $\xi$ and $0.001\hbar/\mu$, respectively.

\begin{figure}[t]
\begin{center}
\includegraphics[keepaspectratio, width=8.6cm,clip]{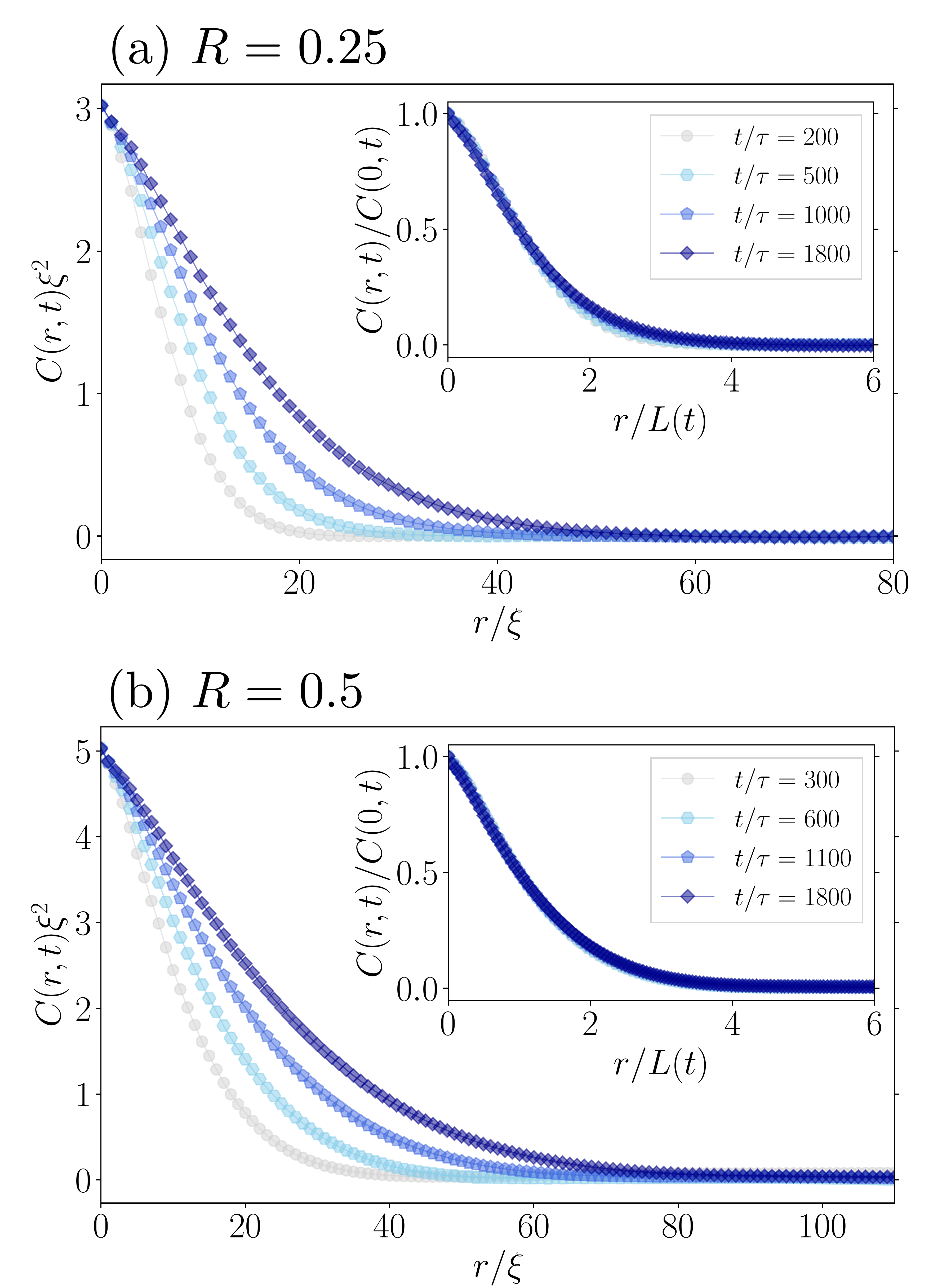}
\caption{(Color online) Time evolution of the correlation functions $C(r,t)$ for $R =  0.25$ (a) and $ 0.5$ (b). The main panels are the raw numerical data showing that the spatial correlations for both cases grow in time. The correlation functions are obtained by taking ensemble averages over $100$ samples with different initial noises generated by Eq.~\eqref{GP9}. (Insets) Correlation functions in the ordinate and abscissa normalized by the on-site correlations $C(0,t)$ and the correlation lengths $L(t)$. The graphs clearly exhibit a signature of the dynamical scaling of Eq.~\eqref{correlation5}. This means that the relaxation dynamics with the large particle-number imbalance even shows the scale-invariant property. \label{fig2} }
\end{center}
\end{figure}

\begin{figure}[t]
\begin{center}
\includegraphics[keepaspectratio, width=8.6cm,clip]{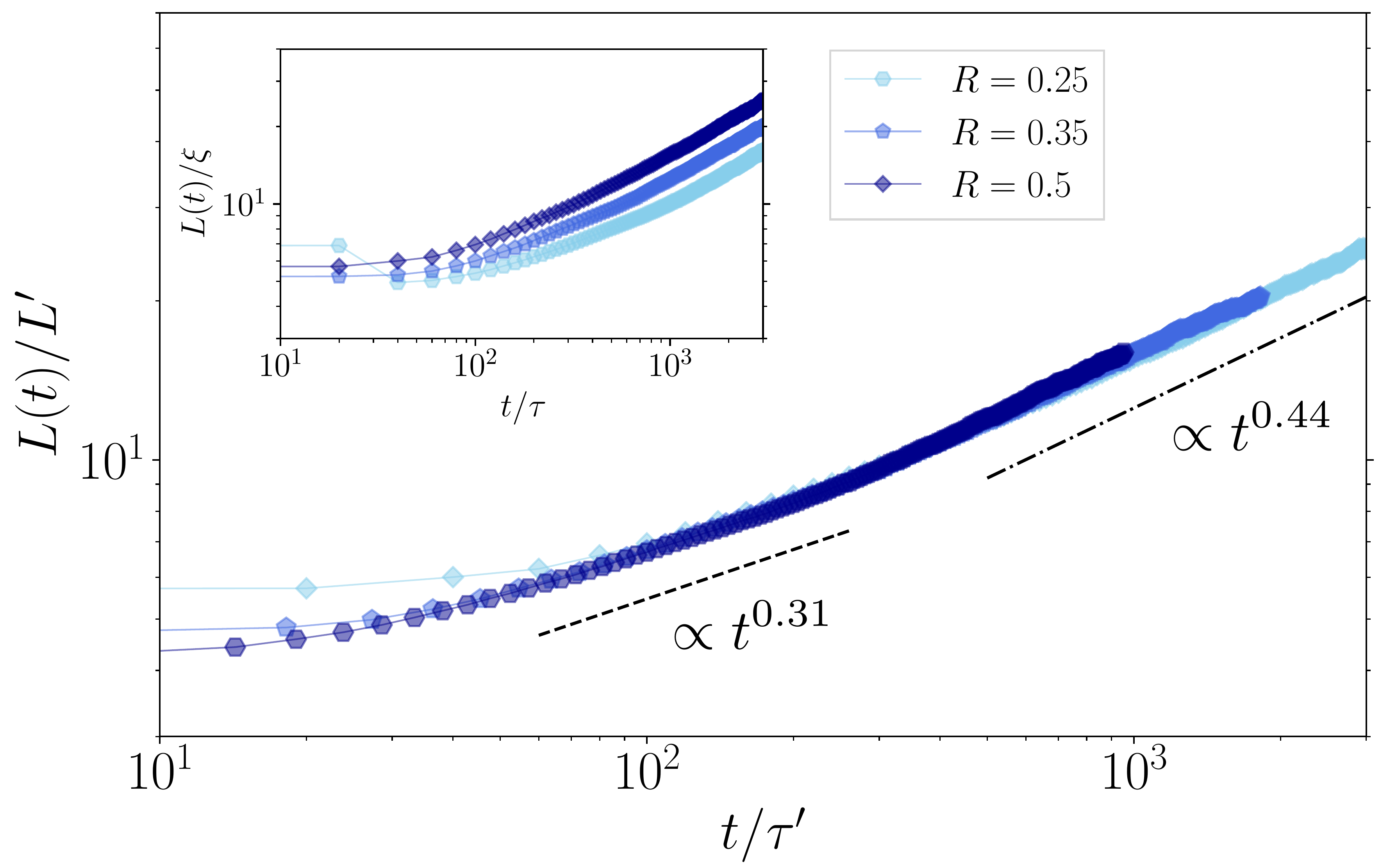}
\caption{(Color online) Time evolution of the correlation lengths $L(t)$ for $R =0.25$, $0.35$, and $0.5$. The inset shows the raw numerical results whereas the main panel plots the same data in the scaled axes. When we normalize the ordinate and abscissa by constants $L'$ and $\tau'$ respectively, the correlation lengths $L(t)$ for different $R$ are almost collapsed into a single curve. Here, we use $(L',\tau')=(1.14 \xi, 4.2 \tau)$, $(1.10 \xi, 2.2 \tau)$, and $(\xi,\tau)$ for $R = 0.25$, $0.35$, and $0.5$. This implies that the relaxation dynamics is identical to each other irrespective of the parameter $R$. The curve has an exponent $\beta \sim 0.31$ in the early stage of the dynamics and it approaches to $0.44$ in the late time. \label{fig3} }
\end{center}
\end{figure}

Figures~\ref{fig1}(c) and~\ref{fig1}(d) show time evolution of $\rho_{2}(\bm{r},t)=|\psi_2(\bm{r},t)|^2$ for $R=0.25$ and $0.5$ after the tuning protocol of Fig.~\ref{fig1}(b). We find that many droplets are generated just after adjusting a target value of $R$. The droplet-nucleation dynamics can be well understood by dynamical instability predicted by the Bogoliubov theory (linear analysis for small fluctuations), from which the most unstable wavelength is given by
\begin{eqnarray}
\frac{\lambda_{\rm in}}{\xi}  =  2 \pi \sqrt{ \frac{2 \mu}{ - \mu +  \sqrt{ g_0^2( \rho_1 - \rho_2)^2 + 4 \rho_1 \rho_2 g_1^2 } } } \label{wavelength} 
\end{eqnarray} 
with $\rho_1 = \rho_{\rm tot} / (1+R)$ and $\rho_2 = \rho_{\rm tot} R/(1+R)$. 
The derivation of Eq.~\eqref{wavelength} is given in Appendix~\ref{instability}. 
Our calculations have $\lambda_{\rm in}/\xi \sim 15.2$ and $13.2$ for $R=0.25$ and $0.5$, respectively, which are consistent with the distributions in the left panels of Figs.~\ref{fig1} (c) and~\ref{fig1}(d). In the late stage of the dynamics, the linear size of each droplet obviously increases in time, and tiny droplet configurations are kept for a longer time when the value of $R$ is smaller. However, as time goes by, such structures are gradually lost, and minor domains become large with fine surface fluctuations. 
Also, one finds a density hole in the domains, which is denoted by a solid arrow in the distributions at $t/\tau=800$ and $1600$ of Fig.~\ref{fig1}(d). 
This structure is a quantized vortex for the component 2, and there the component 1 occupies the vortex core. 
Because of topological stability, the composite vortex survives until when it encounters its anti-vortex or goes outside of the domain. However, such a small structure is expected to hardly affect the domain growth dynamics.

To see the scale-invariant aspects of the relaxation dynamics, we calculate the correlation function \eqref{correlation1}, and find the dynamical scaling of Eq.~\eqref{correlation5} as shown in Fig.~\ref{fig2}. The main panels of Fig.~\ref{fig2} are the plots of the raw numerical data, which clearly exhibit monotonous growth of the correlation lengths in time for both cases. The insets show the same data in the normalized ordinate and abscissa by the on-site correlations $C(0,t)$ and the correlation lengths $L(t)$ respectively, clearly demonstrating that the numerical data at the different times can collapse to a single curve. This behavior is consistent with the dynamical scaling of Eq.~\eqref{correlation5}, which is a hallmark of scale-invariant relaxation dynamics.

The dynamical scaling is characterized by the exponents $\beta$ and $\gamma$. To determine $\beta$, we first plot the time evolution of $L(t)$. The inset of Fig.~\ref{fig3} shows the time evolution of $L(t)$ for $R=0.25$, $0.35$, and $0.5$. These curves seemingly show different time dependences, but they can be expressed by a single function when we divide the ordinate and abscissa for each $R$ by constants $L'$ and $\tau '$, respectively, as shown in the main panel of Fig.~\ref{fig3}. This implies that the relaxation dynamics for the different imbalance parameters is identical to each other. From the figure, we find that the exponent $\beta$ is close to $0.31$ in the early time and approaches to $0.44$ in the late. These exponents are estimated by fitting a power law function $y=ax^b$ to the data with the Marquardt-Levenberg method in Gunplot. Here, $a$ and $b$ are the fitting parameters. The fitting regions for the two exponents are determined by eye. On the other hand, as we mentioned in Sec.~\ref{correlation_review}, the exponent $\gamma$ is zero because the particle number $N_2$ is conserved in time after the tuning protocol and $C(0,t)$ is identical to $N_2/S$. Thus, we obtain the following scaling exponents:
\begin{eqnarray}
(\beta,  \gamma ) \simeq
\begin{cases}
\displaystyle
 (0.31, 0) & (t < t^{*}),  \\
 \displaystyle
 (0.44, 0) & (t^{*}<t)
\end{cases}
\end{eqnarray} 
with the crossover time $t^*$. As described in the next section, the fundamental mechanism behind the crossover is that the droplets randomly move with the droplet-size-dependent velocity.

We also investigate the density correlation function in the relaxation dynamics, and find that the correlation function does not obey the dynamical scaling when $R$ is far from unity. We expect that this is attributed to existence of two length scales: one is a typical domain size and the other is an average distance between droplets. As mentioned in Sec.~\ref{correlation_review}, the density correlation function includes contributions from a density distribution around a droplet, which related to the averaged distance. Thus, the correlation function cannot be described by the dynamical scaling because Eq.~\eqref{correlation5} means that the system is characterized only by a single length scale.

Before closing this section, we should note that the interaction-parameters in the simulations are far from the transition points ($g_0=g_1$). In this sense
our result of the relaxation dynamics supports the NTFP conjecture described in Sec.~\ref{correlation_review}.

\section{Analytical derivation of the power laws for the correlation length \label{ana_derivation}}
We analytically derive power laws of the correlation length $L(t)$ using the GP equation \eqref{GP1} and \eqref{GP2}. The essential idea is that many droplets of the minor component randomly move with the size-dependent velocity $V( L_{\rm dp}(t) )$, where $L_{\rm dp}(t)$ is the linear size of a droplet. The derivation is divided into two steps. Firstly, focusing on the dynamics of a single droplet, we show that the velocity obeys two power laws $V(L_{\rm dp}) \propto L_{\rm dp}^{-n}~(n=1,2)$. Secondly, assuming random collisions between droplets, we construct an equation of motion for the averaged droplet size. As a result, we derive the power exponent $\beta=1/3$ and $1/2$ for the early and late stages of the dynamics, respectively. Our theoretical results well describe the numerical results presented in Fig.~\ref{fig3}.

\subsection{Dynamics of a single droplet}
We investigate dynamics of a single droplet by employing a continuity equation of spin density vectors defined by 
\begin{eqnarray}
f_{\nu} = \frac{1}{\rho }\sum_{m,n=1,2} \psi_{m}^{*} (\sigma_{\nu})_{mn} \psi_{n}, 
\end{eqnarray} 
\begin{eqnarray}
\rho = |\psi_1|^2 + |\psi_2|^2
\end{eqnarray} 
with the Pauli matrix $(\sigma_{\nu})_{mn}~(\nu=x,y,z)$. The continuity equation for $f_z$ \cite{hydro0,hydro1,hydro2,hydro3} is given by
\begin{eqnarray}
\frac{\partial}{\partial t} \rho f_{z} + {\bm \nabla} \cdot {\bm J}_{z} = 0, \label{spin1} 
\end{eqnarray} 
\begin{eqnarray}
{\bm J}_{z} = \rho \biggl[ f_z \bm{v} + \frac{\hbar}{2M}  \Big( f_y {\bm \nabla} f_x - f_x {\bm \nabla} f_y  \Big)\biggl], \label{spin2} 
\end{eqnarray} 
where ${\bm J}_z$ is the spin current for $f_z$ and $\bm{v}$ being the mass current  defined by
\begin{eqnarray}
\bm{v} = \frac{\hbar}{2M\rho i} \sum_{m=1}^{2}  \Bigl( \psi_{m}^* \bm{\nabla} \psi_{m} - \psi_{m} \bm{\nabla} \psi_{m}^* \Bigl). \label{spin4}
\end{eqnarray} 
The derivation of Eqs.~\eqref{spin1} -- \eqref{spin4} is explained in Appendix~\ref{fz_con}. It is worthy to note that the total density distribution $\rho$ is almost uniform because of a small repulsion interaction $ |g_0-g_1|/(g_0+g_1) = 0.2$ between the different components. In the following, we assume $\rho \simeq \rho_{\rm tot}$. 

\begin{figure}[t]
\begin{center}
\includegraphics[keepaspectratio, width=8.6cm,clip]{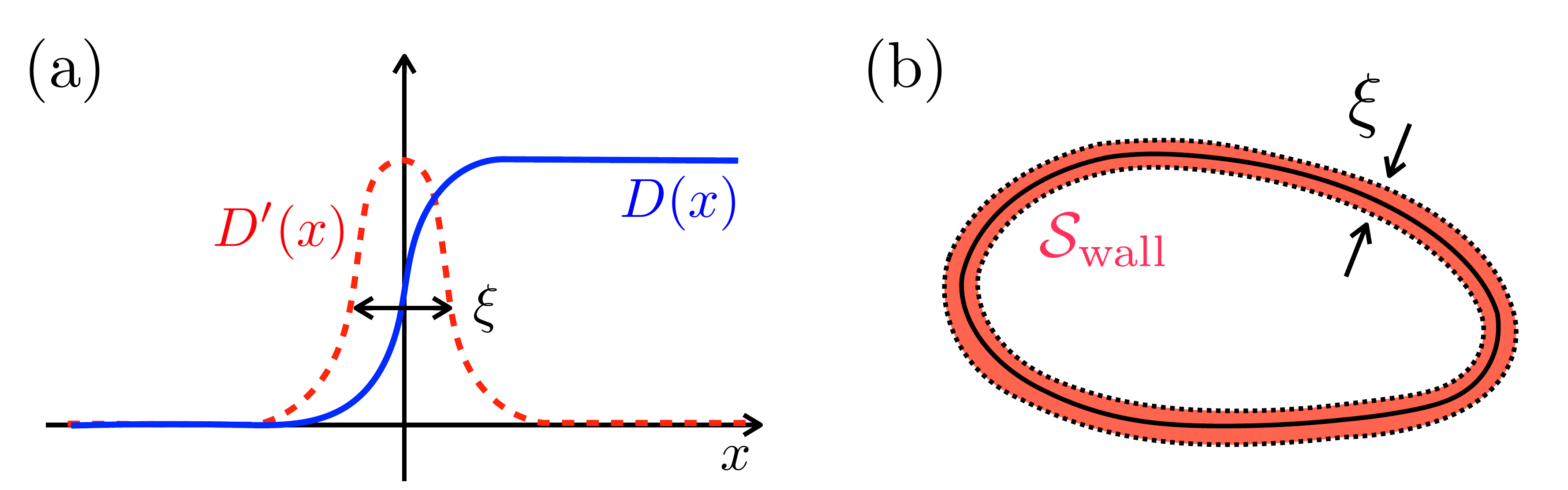}
\caption{(Color online) (a) Schematic for the domain configuration $D(x)$ (solid line) and it's derivative $D'(x)$ (dashed line). The typical width is in the order of the coherence length $\xi$. (b) Integral region $\mathcal{S}_{\rm wall}$. The solid lines depict the edge of the droplet, around which the spatial derivative of $D(x)$ has a large value. The region surrounded by the dashed lines is $\mathcal{S}_{\rm wall}$, which gives main contribution to the spatial integral of Eq.~\eqref{droplet4}.
\label{fig4} }
\end{center}
\end{figure}

Here, we assume that there is a single droplet of the minor component in the system. The centroid ${\bm X}(t)$ of the droplet is defined by
\begin{eqnarray}
\bm{X}(t) = \frac{   \int_{\mathcal{S}} \bm{r} D(\bm{r},t) d\bm{r}   }{  \int_{\mathcal{S}} D(\bm{r},t)  d\bm{r}},   \label{droplet1} 
\end{eqnarray} 
\begin{eqnarray}
D(\bm{r},t) = \frac{1}{2} \Bigl( 1- f_z(\bm{r},t) \Bigl).   \label{droplet1_1} 
\end{eqnarray} 
The function $D(x)$ describes a reshaped droplet configuration as shown in Fig.~\ref{fig4}(a), which has a sharp edge with a width in the order of $\xi$.

\begin{figure}[t]
\begin{center}
\includegraphics[keepaspectratio, width=8.6cm,clip]{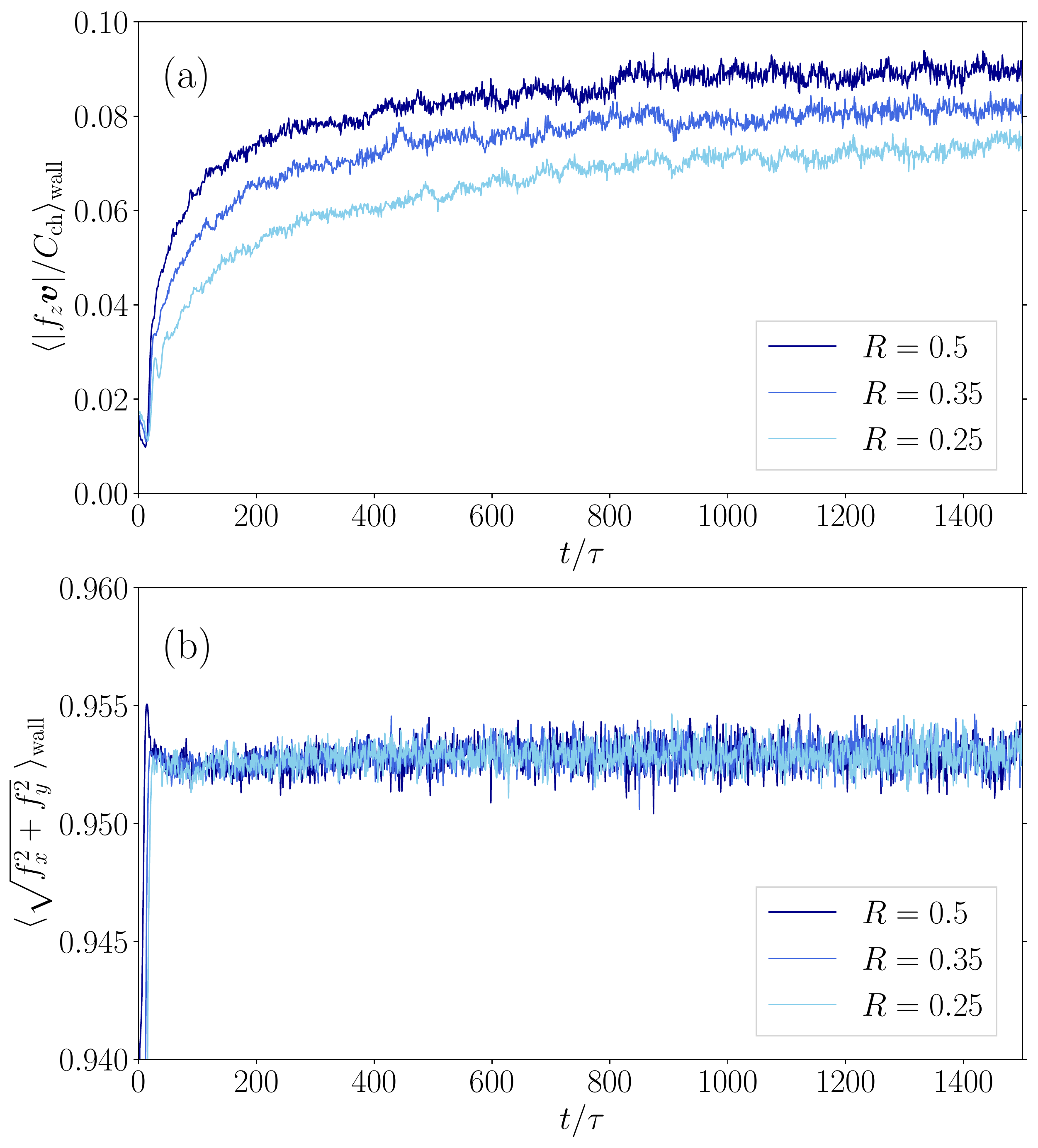}
\caption{(Color online) Time evolution of (a) $\langle |f_z \bm{v}| / C_{\rm ch} \rangle_{\rm wall}$ and (b) $\langle \sqrt{f_x^2 + f_y^2} ~\rangle_{\rm wall}$ for a single sample corresponding to Fig.~\ref{fig1}(c). Here, the bracket $\langle \cdots \rangle_{\rm wall}$ denotes a spatial average around the domain wall defined by $\{ \bm{r} |  -0.5 < f_z(\bm{r})  < 0.5  \}$. The upper panel (a) shows that the velocity becomes smaller when the imbalance parameter $R$ is smaller. This reflects the fact that the instability predicted by the Bogoliubov theory is weaker for the smaller $R$ as described in Appendix~\ref{instability}.
On the other hand, the lower panel (b) shows that the amplitude of the transverse spin is almost unity irrespective of $R$ because $f_z$ becomes small in $\mathcal{S}_{\rm wall}$.
\label{fig5} }
\end{center}
\end{figure}

We then calculate the time derivative of ${\bm X}(t)$. We start with a time derivative of the denominator of Eq.~\eqref{droplet1} from the perspective of stability of the single droplet. By definition, the denominator is almost the same as the droplet area $S_{\rm dp}$ because the droplet has the flat profile of $f_z$ except for the edge:
\begin{eqnarray}
S_{\rm dp} \simeq \int_{\mathcal{S}} D( \bm{r},t )  d\bm{r}.   \label{droplet2} 
\end{eqnarray} 
Since our model is symmetric under rotation about the $z$-axis in the spin space, the net longitudinal magnetization should be conserved:
\begin{eqnarray}
\int_{\mathcal{S}} f_z( \bm{r},t )  d\bm{r} = {\rm Const}.
\end{eqnarray} 
Thus, we can safely assume that the area $S_{\rm dp}$ is independent of time. This means that the single droplet itself is quite stable and the area is conserved although the shape can change in time.

Utilizing Eqs.~\eqref{spin1} and \eqref{droplet2}, we calculate the time derivative of $X_{\nu}(t)~(\nu=x,y)$:
\begin{eqnarray}
\frac{d}{dt}X_{\nu}(t) &\simeq& \frac{1}{S_{\rm dp}} \frac{d}{d t} \int_{\mathcal{S}} r_{\nu} D( \bm{r},t) d\bm{r}, \nonumber \\ 
&=& -\frac{1}{ 2S_{\rm dp} }  \int_{\mathcal{S}} r_{\nu} \frac{\partial}{\partial t} f_{z}(\bm{r},t)  d\bm{r}, \nonumber \\ 
&=&  \frac{1}{2\rho_{\rm tot} S_{\rm dp} }  \int_{\mathcal{S}} r_{\nu} \bm{\nabla} \cdot \bm{J}_{z}(\bm{r},t)  d\bm{r}. \nonumber \\ 
\label{droplet4} 
\end{eqnarray} 
Here, note that the divergence $\bm{\nabla} \cdot \bm{J}_{z}(\bm{r},t)$ should have large values around the center of the domain wall because $D(\bm{r})$ rapidly changes over the edge as shown in Fig.~\ref{fig4}(a). The typical length scale for this change is $\xi$, and the main contribution of the spatial integral to Eq.~\eqref{droplet4} is expected to come from $\mathcal{S}_{\rm wall}$ schematically shown in Fig.~\ref{fig4}(b). Thus, we can replace the integral region $\mathcal{S}$ by $\mathcal{S}_{\rm wall}$: 
\begin{eqnarray}
\frac{d}{dt}X_{\nu}(t) \simeq \frac{1}{ 2\rho_{\rm tot} S_{\rm dp} }  \int_{\mathcal{S}_{\rm wall}} r_{\nu} \bm{\nabla} \cdot \bm{J}_{z}(\bm{r},t)  d\bm{r}.
\label{droplet5} 
\end{eqnarray} 

To analyze the integral \eqref{droplet5}, we evaluate the spin current $\bm{J}_{z}(\bm{r},t)$ of Eq.~\eqref{spin2}. We normalize it by the characteristic length $\xi$ and time $\tau$, and the current becomes
\begin{eqnarray}
\frac{ {\bm J}_{z}}{ \rho_{\rm tot} C_{\rm ch} } = \frac{f_z \bm{v}}{C_{\rm ch}} + f_y  \bar{{\bm \nabla}} f_x - f_x  \bar{{\bm \nabla}} f_y
\label{droplet6} 
\end{eqnarray} 
with the non-dimensional gradient $\bar{\bm \nabla} = \xi \bm{\nabla}$ and the characteristic velocity $C_{\rm ch} = \xi / \tau$. 
To estimate the order of each term on the right-hand side of Eq.~\eqref{droplet6}, we assume that the spatial gradients in Eq.~\eqref{droplet6} can be  replaced by the size of the droplet $L_{\rm dp}$ because of the following reason. In the relaxation dynamics in Figs.~\ref{fig1}(c) and (d), surfaces of droplets have various excitations whose wavelengths are smaller than $L_{\rm dp}$. However, we expect that such fine surface excitations are not essential in the droplet dynamics since the effect may be canceled in the integral of Eq.~\eqref{droplet5} over space. 
Thus, the main contribution to the integral is expected to come from the spatial structures of $f_{\nu}(\bm{r},t)$ and $v_{\nu}(\bm{r},t)$ on the scale of the droplet size $L_{\rm dp}$.
By using this approximation, we can obtain the following naive condition to neglect the mass current term in Eq.~\eqref{droplet6}:
\begin{eqnarray}
\biggl\langle  \frac{|f_z \bm{v}|}{C_{\rm ch}}   \biggl\rangle_{\rm wall} &<&   \langle f_x^2 + f_y^2  \rangle_{\rm wall}  \frac{\xi}{L_{\rm dp}}  \label{droplet7_0}  \\
\nonumber \\
\rightarrow L_{\rm dp} &\lesssim&  \Lambda\xi, \\
\nonumber 
\label{droplet7} 
\end{eqnarray} 
\begin{eqnarray}
 \Lambda = \frac{ C_{\rm ch}  \langle f_x^2+f_y^2  \rangle_{\rm wall} }{  \langle |f_z \bm{v}|  \rangle_{\rm wall} }, 
 \label{droplet7_1} 
\end{eqnarray} 
where $\langle \cdots \rangle_{\rm wall}$ the spatial average inside the wall $\mathcal{S}_{\rm wall}$.
The left (right)-hand side on inequality~\eqref{droplet7_0} comes from the first (second) term of the right-hand side on Eq.~\eqref{droplet6}.

To obtain the values of $\Lambda$, we numerically calculate the absolute values of $f_z {\bm v}_{\nu}/C_{\rm ch}$ and $f_{\nu}$ and average them inside the walls. As shown in Fig.~\ref{fig5}, $\sqrt{ f_{x}^2 + f_{y}^2}$ is about $0.953$ irrespective of $R$ while $|f_z \bm{v}/C_{\rm ch}|$ is about $0.07$, $0.08$, and $0.09$ for $R=0.25$, $0.35$, and $0.5$, respectively. Thus, our numerical simulations for $R=0.25$, $0.35$, and $0.5$ leads to $\Lambda = 13.0$, $11.4$, and  $10.1$, respectively. 
Combining all the results obtained so far, we derive an expression for the spin current depending on the droplet size $L_{\rm dp}$:   
\begin{eqnarray}
{\bm J}_{z} \simeq 
\begin{cases}
\displaystyle
\frac{\rho_{\rm tot}\hbar}{2M}  \Big( f_y {\bm \nabla} f_x - f_x {\bm \nabla} f_y  \Big) & (L_{\rm dp} \lesssim \Lambda \xi), \label{spin3}  \\
\displaystyle
\rho_{\rm tot} f_z \bm{v} & (L_{\rm dp} \gtrsim \Lambda \xi).
\end{cases}
\label{spin_current_change}
\end{eqnarray} 

Substituting Eq.~\eqref{spin3} into Eq.~\eqref{droplet5}, we obtain an expression for the centroid velocity $X_{\nu}(t)~(\nu=x,y)$ of the single droplet:
\begin{widetext}
\begin{eqnarray}
\frac{d}{dt}X_{\nu}(t) \simeq
\begin{cases}
\displaystyle
 \frac{\hbar}{ 4M S_{\rm dp} }  \int_{\mathcal{S}_{\rm wall}} r_{\nu}  \Big( f_y(\bm{r},t) {\bm \nabla}^2 f_x(\bm{r},t) - f_x(\bm{r},t) {\bm \nabla}^2 f_y(\bm{r},t)  \Big)  d\bm{r} & (L_{\rm dp} \lesssim \Lambda \xi), \\
 \displaystyle
 \frac{1}{ 2S_{\rm dp} }  \int_{\mathcal{S}_{\rm wall}} r_{\nu}  \bm{\nabla} \cdot \big(  \bm{v}(\bm{r},t) f_z(\bm{r},t)  \big)  d\bm{r}  & (L_{\rm dp} \gtrsim \Lambda \xi).
\end{cases}
\label{droplet8} 
\end{eqnarray}
\end{widetext}
This demonstrates that the main driving mechanisms for the droplet can change in accordance with the droplet size. As described in the next subsection, the change of the predominant term leads to the two power laws, and thus we can predict the crossover phenomena of the scaling in Fig.~\ref{fig3}. 

Finally, we estimate the $L_{\rm dp}$dependence of the integral in the right-hand side of Eq.~\eqref{droplet8}. We notice that the spin density $f_{\nu}$ and the mass current $\bm{v}$ in $\mathcal{S}_{\rm wall}$ are expected to modulate on the scale of $L_{\rm dp}$, and thus the integral can be estimated as  
\begin{widetext}
\begin{eqnarray}
\int_{\mathcal{S}_{\rm wall}} r_{\nu}  \Big( f_y(\bm{r},t) {\bm \nabla}^2 f_x(\bm{r},t) - f_x(\bm{r},t) {\bm \nabla}^2 f_y(\bm{r},t)  \Big)  d\bm{r} \sim L_{\rm dp}^2 \times \frac{1}{L_{\rm dp}^2} \sim \mathcal{O}(1),
\label{droplet9_1} 
\end{eqnarray} 
\begin{eqnarray}
 \int_{\mathcal{S}_{\rm wall}} r_{\nu}   \bm{\nabla} \cdot \big(  \bm{v}(\bm{r},t) f_z(\bm{r},t) \big) d\bm{r}  &\sim& L_{\rm dp}^2 \times \frac{1}{L_{\rm dp}} \sim \mathcal{O}(L_{\rm dp}).
\label{droplet9_2} 
\end{eqnarray} 
\end{widetext}

$\\$
$\\$
Combining Eqs.~\eqref{droplet8}, \eqref{droplet9_1} and \eqref{droplet9_2}, we conclude
\begin{eqnarray}
V(L_{\rm dp}) = \frac{d}{dt}X_{\nu}(t) \propto
\begin{cases}
\displaystyle
 \frac{1}{L_{\rm dp}^2} & (L_{\rm dp} \lesssim \Lambda \xi), \\
\displaystyle
  \frac{1}{L_{\rm dp}} & (L_{\rm dp} \gtrsim \Lambda \xi).
\end{cases}
\label{droplet10} 
\end{eqnarray} 
Here, we use  the relation $S_{\rm dp} \propto L_{\rm dp}^2$.

\begin{table}[b]
      \begin{tabular}{cp{0.5cm}cp{0.5cm}cp{0.5cm}c}
        $R$  & & $\Lambda$ &  &  $\Lambda(R)/\Lambda(0.5)$ &  & $L'/\xi$   \\[3pt] \hline\hline
        0.5   & & 10.1            &  & 1                                &  &  1            \\ [3pt]
        0.35 & & 11.4            &  & 1.13                           &  &   1.10       \\ [3pt]
        0.25 & & 13.0            &  & 1.29                           &  &   1.14       \\ [3pt] \hline
      \end{tabular}
      \caption{Dependence of $\Lambda$ and $L'/\xi$ in Fig.~\ref{fig3} on the imbalance parameter $R$. 
      The third column shows $\Lambda$ for $R=0.25$ and $0.35$ divided by that for $R=0.5$. Despite the naive estimation of $\Lambda$, the values in the third column are reasonably close to those in the fourth one.}
      \label{table1}
\end{table}

\subsection{Growth law of droplets \label{grwoth_law}}
We construct the equation of motion for the correlation length $L(t)$ using Eq.~\eqref{droplet10}, and derive the $1/3$ and $1/2$ power laws.
Suppose that we have many droplets with the average size $L_{\rm dp}(t)$ and the mean distance between them is denoted by $L_{\rm md}(t)$. 
Then, the number $N_{\rm dp}(t)$ of the droplets is given by 
\begin{eqnarray}
N_{\rm dp}(t) = \frac{A}{L_{\rm md}(t)^2}
\label{kinetic1} 
\end{eqnarray} 
with a constant $A$ proportional to the system area $S$.
Assuming that these droplets merge through random collisions between them, we expect that an average collision-time is about $L_{\rm md}(t)/ V(L_{\rm dp}(t))$. Thus, the the simple rate equation for $N_{\rm dp}(t)$ is given by
\begin{eqnarray}
\frac{d}{dt} N_{\rm dp}(t) = - B \frac{ V(L_{\rm dp}(t)) } { L_{\rm md}(t) } N_{\rm dp}(t)
\label{kinetic2} 
\end{eqnarray} 
with a constant $B$. We note that $L_{\rm md}(t)$ is proportional to $L_{\rm dp}(t)$:  
\begin{eqnarray}
L_{\rm md}(t)  \propto L_{\rm dp}(t).
\label{kinetic3} 
\end{eqnarray} 
This relation can be derived by considering the conservation laws for $N_1$ and $N_2$, the detail of which is described in the Appendix~\ref{proportion}. 
Substituting Eqs.~\eqref{droplet10},~\eqref{kinetic1}, and \eqref{kinetic3} into Eq.~\eqref{kinetic2}, we obtain 
\begin{eqnarray}
\frac{d}{dt} L_{\rm dp}(t) 
\propto 
\begin{cases}
\displaystyle L_{\rm dp}(t)^{-2} & (L_{\rm dp} \lesssim \Lambda \xi),  \\
\displaystyle L_{\rm dp}(t)^{-1} & (L_{\rm dp} \gtrsim \Lambda \xi).
\end{cases}
\label{kinetic4} 
\end{eqnarray} 
Regarding that the droplet size $L_{\rm dp}(t)$ is proportional to the correlation length $L(t)$, we finally obtain the two power laws by solving Eq.~\eqref{kinetic4}: 
\begin{eqnarray}
L(t) \propto 
\begin{cases}
\displaystyle t^{1/3} & (L(t) \lesssim \Lambda \xi),  \\
\displaystyle t^{1/2} & (L(t) \gtrsim \Lambda \xi).
\end{cases}
\label{kinetic5} 
\end{eqnarray} 
This shows that the correlation length monotonically grows in time, and thus we readily expect that the $1/3$ power law appears in the early stage of the dynamics and the $1/2$ power law gradually does in the late time.

Comparing Eq.~\eqref{kinetic5} with the numerical results of Fig.~\ref{fig3}, we notice that there is a small difference for the exponent in the early time, but the entire evolution of $L(t)$ is consistent with our theoretical prediction. To quantify the consistency, we compare the values of the relation between $L'$ used in Fig,~\ref{fig3} and $\Lambda \xi$. The former is the rescaling factor of $L(t)$ which is defined such that the curves for $R=0.25$ and $0.35$ coincide with that for $R=0.5$. In other words, $L'/\xi$'s for $R=0.25$ and $0.35$ give the ratio of the crossover lengths $\Lambda \xi$, at which the power law changes, to that for $R=0.5$. Thus, the values of $\Lambda \xi$ for $R=0.25$ and $0.35$ divided by that for $R=0.5$ can be close to $L'/\xi$ because $\Lambda \xi$ is a measure of the crossover points. We summarize the numerical numbers in Table~\ref{table1}, which shows that the two quantities have the reasonable values despite the naive estimation of $\Lambda$ in Eq.~\eqref{droplet7_1}.

\section{Discussion \label{discussion}}
We discuss four issues. Firstly, we consider stability of droplet structures, which is the assumption used in our analytical model of Sec.~\ref{ana_derivation}. Secondly, introducing previous works on relaxation dynamics with large imbalances in classical systems, we clarify the difference and argue that the mechanism of the power laws in our system is unconventional. 
Thirdly, the dependence of the relaxation dynamics on the imbalance parameter $R$ is discussed. Finally, we propose an experimental setup to observe our theoretical prediction such as crossover phenomena.

\subsection{Stability of isolated droplets \label{SD}}
Our analytical model described in Sec.~\ref{ana_derivation} has the important assumption that an isolated droplet is stable against fluctuations behind it. Here, we verify this by numerically investigating the isolated droplet dynamics.   

\subsubsection{Droplet dynamics in a realistic situation}
We prepare an initial state with four isolated droplets using the imaginary time step of Eqs.~\eqref{GP1} and \eqref{GP2} with $\Omega=0$.
To stabilize the droplets with the sizes $L_{p}/\xi = 8,10,12$, and $14$, we apply the following external potential $V_{p}$ only to the component $1$:
\begin{eqnarray}
V_{p}(\bm{r}) = 
\begin{cases}
\displaystyle
2 \mu & ( \bm{r} \in  \mathcal{A}_{\rm dp}); \\
\displaystyle
0 & ({\rm otherwise}),
\end{cases}
\label{potential1} 
\end{eqnarray} 
where the bulk chemical potential $\mu$ is identical to that in Figs.~\ref{fig1}(c) and (d). 
Here, the region $\mathcal{A}_{\rm dp}$ is given by $\mathcal{A}_1 \cup \mathcal{A}_2 \cup \mathcal{A}_3 \cup \mathcal{A}_4$ with
\begin{eqnarray}
\mathcal{A}_{j} = \{ \bm{r} |  (x-x_{j})^2 + (y-y_{j})^2 < R_{j}^2 \},
\label{potential2} 
\end{eqnarray} 
where the potential position $(x_j, y_j)$ and the radius $R_{j}$. In our simulation, we use $(x_1/\xi,y_1/\xi,R_1/\xi)=(64,64,4)$, $(x_2/\xi,y_2/\xi,R_2/\xi)=(192,64,5)$, $(x_3/\xi,y_3/\xi,R_3/\xi)=(64,192,6)$, and $(x_4/\xi,y_4/\xi,R_4/\xi)=(192,192,7)$. 
As a result, we obtain a macroscopic wavefunction $\psi_{m}^{\rm (dp)}(\bm{r})$ with the four droplets. 

We add initial noises into $\psi_{m}^{\rm (dp)}(\bm{r})$ to mimic droplet motions in the quench dynamics of Figs.~\ref{fig1}(c) and (d), and then numerically solve Eqs.~\eqref{GP1} and \eqref{GP2} under $\Omega=0$ and $V_{p}=0$. The initial wavefunction used here is
\begin{eqnarray}
\psi_1(\bm{r},0) &=& \psi_{1}^{\rm (dp)}(\bm{r}) \nonumber \\
&+& \frac{1}{\sqrt{S}} \sum_{\bm{k} \in \{ \bm{k} | E_{\bm{k}} \leq \mu, \bm{k} \neq 0 \} } \bigl( u_{\bm{k}}\beta_{1,\bm{k}} + v_{\bm{k}}\beta_{1,-\bm{k}}^{*}  \bigl) e^{i\bm{k}\cdot\bm{r}},  \nonumber \\
\label{noise2_1}
\end{eqnarray}
\begin{eqnarray}
\psi_2(\bm{r},0) = \psi_{2}^{\rm (dp)}(\bm{r}) + \frac{1}{\sqrt{S}} \sum_{\bm{k} \in \{ \bm{k} | \epsilon_{\bm{k}} \leq \mu, \bm{k} \neq 0 \}} \beta_{2,\bm{k}} e^{i\bm{k}\cdot\bm{r}}, \label{noise2}
\end{eqnarray}
where $\beta_{1,\bm{k}}$ and $\beta_{2,\bm{k}}$ are sampled from the probability distribution functions defined by
\begin{eqnarray}
W_1( \{ \beta_{1,\bm{k}} \}) = B_1 \prod_{\bm{k}} {\rm exp} \left( -  \frac{ | \beta_{1,\bm{k}} |^2 }{2b_1^2}  \right), 
\end{eqnarray}
\begin{eqnarray}
W_2( \{ \beta_{2,\bm{k}} \}) = B_2 \prod_{\bm{k}} {\rm exp} \left( - \frac{ | \beta_{2,\bm{k}} |^2 }{ 2b_2^2 } \right).
\end{eqnarray}
Here $B_1$ and $B_2$ are normalization constants, and widths $b_1$ and $b_2$ of the distributions are $2.7$ and $0.6$, which give almost same amplitude of the density fluctuations after the domain nucleation in Fig.~\ref{fig1}. The form of these noises are identical to Eqs.~\eqref{GP5_1} and \eqref{GP5}, except for the values of $b_1$ and $b_2$.

\begin{figure}[t]
\begin{center}
\includegraphics[keepaspectratio, width=8.6cm,clip]{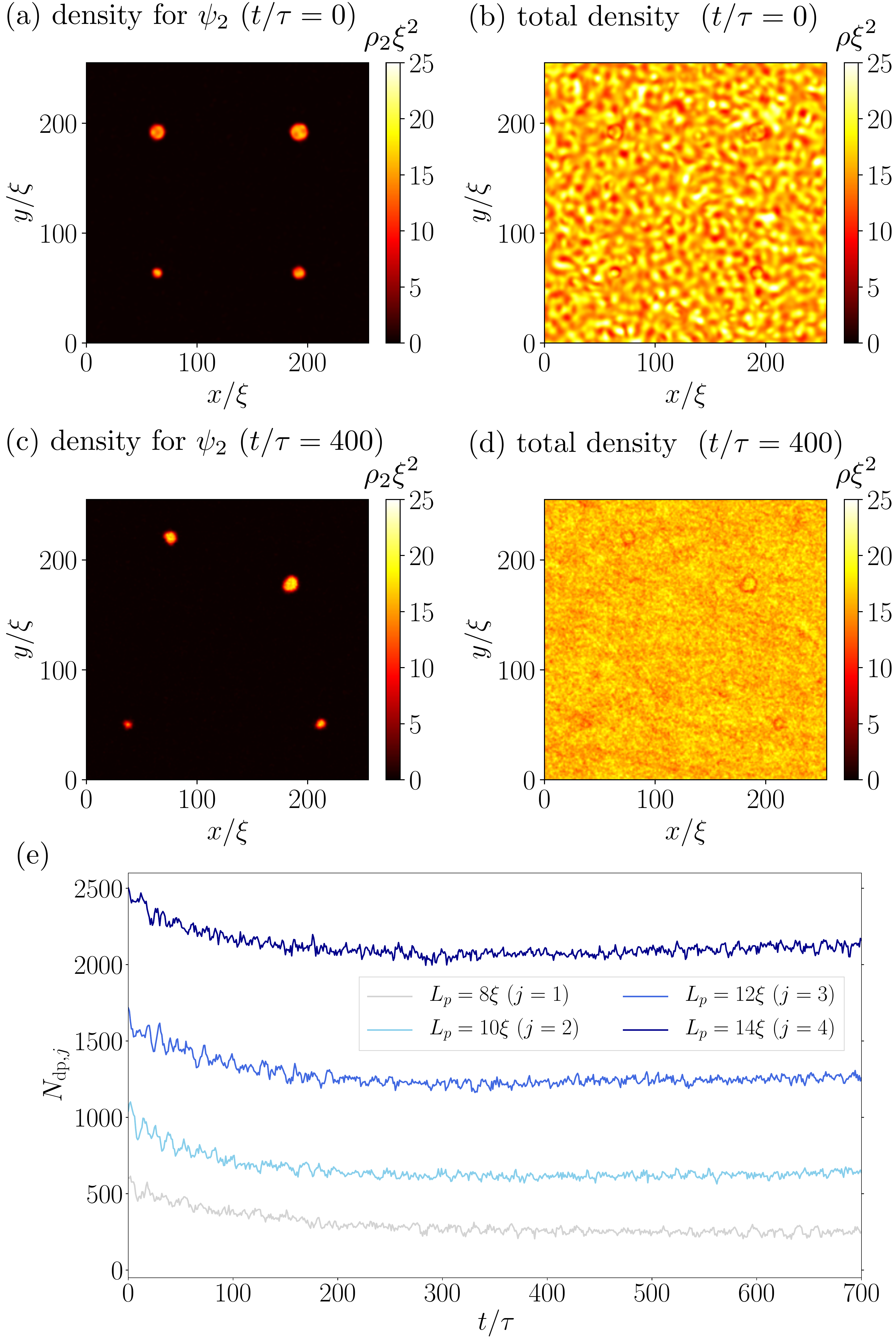}
\caption{
(Color online) Dynamics of four isolated droplets with the sizes $L_{p}/\xi=8,10,12$, and $14$ in a realistic situation similar to the quench dynamics. 
Panels (a) and (b) shows initial distributions for the density $\rho_{2}(\bm{r},t)$ of the component 2 and the total density $\rho(\bm{r},t)$. In the initial state, we add noises to the macroscopic wavefunction in order to mimic the quench dynamics. The time-evolved distributions at $t/\tau=400$ are shown in panels (c) and (d), in which we find that the four droplets survive in the long time dynamics under the fluctuations. Panel (e) shows time evolution of the droplet particle numbers $N_{{\rm dp},j}~{j=1, \ldots ,4}$. This indicates that the particles contained in each droplet is approximately conserved in time and the isolated droplets are stable.  
\label{fig6} }
\end{center}
\end{figure}

Figures~\ref{fig6}(a)--(d) show time evolution for spatial distributions of $\rho_{2}(\bm{r},t)$ and $\rho(\bm{r},t)$, from which we find that all the isolated droplets are stable. 
We quantitatively estimate the stability by calculating a particle number contained in each droplet, which is measured by
\begin{eqnarray}
N_{{\rm dp},j} = \int_{\mathcal B_j(t)} \rho_{2} (\bm{r},t) , 
\end{eqnarray} 
with $\mathcal{B}_j(t) = \{ \bm{r} | \rho_2(\bm{r},t) > \rho_{\rm tot} /2 \} \cap \{ \bm{r} | x_{1,j} \leq  x < x_{2,j}, y_{1,j} \leq y < y_{2,j} \}$. 
Here, the coordinates used here are given by $(x_{1,1},x_{2,1},y_{1,1},y_{2,1}) = (0,128\xi,0,128\xi)$, $(x_{1,2},x_{2,2},y_{1,2},y_{2,2}) = (129\xi,255\xi,0,128\xi)$, $(x_{1,3},x_{2,3},y_{1,3},y_{2,3}) = (0,128\xi,129\xi,255\xi)$, and $(x_{1,4},x_{2,4},y_{1,4},y_{2,4}) = (129\xi,255\xi,129\xi,255\xi)$. 
As shown in Fig.~\ref{fig6}(e), we find that the particle numbers in the droplets are approximately conserved in the long time dynamics. 
Here, we note the early stage of the dynamics, in which there are small decreases of the droplet particle numbers.
We expect that this is caused by instabilities arising from our initial droplets, which are externally stabilized by applying the potential $V_{p}(\bm{r})$. Thus the initially prepared droplets are slightly different from the genuine configurations, and emit several partilces to reduce the sizes. 

\begin{figure}[t]
\begin{center}
\includegraphics[keepaspectratio, width=9cm,clip]{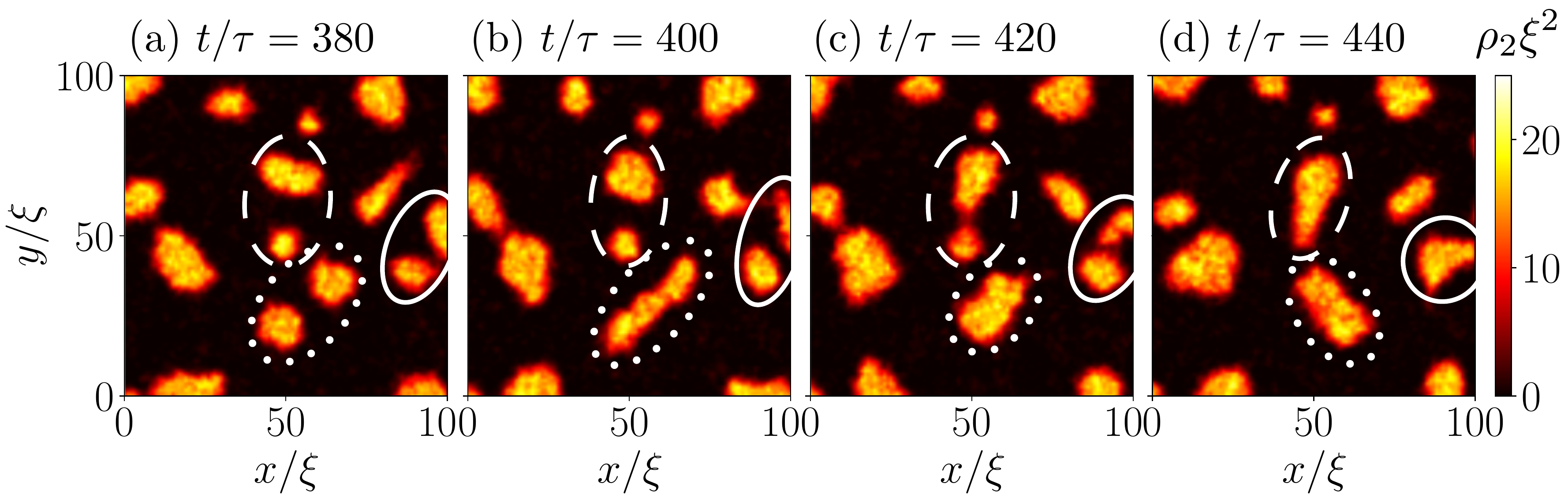}
\caption{
(Color online) Domain merging dynamics of Fig.~\ref{fig1}(c) after the quench. Three merging processes are encircled by solid, dashed, and dotted lines. 
\label{fig6_2} }
\end{center}
\end{figure}

In the simulation of the quench dynamic described in Sec.~\ref{Num}, we indeed confirm that the stable droplets collide each other, and the typical size grows in time as shown in Fig.~\ref{fig6_2}. 

\begin{figure}[t]
\begin{center}
\includegraphics[keepaspectratio, width=8.6cm,clip]{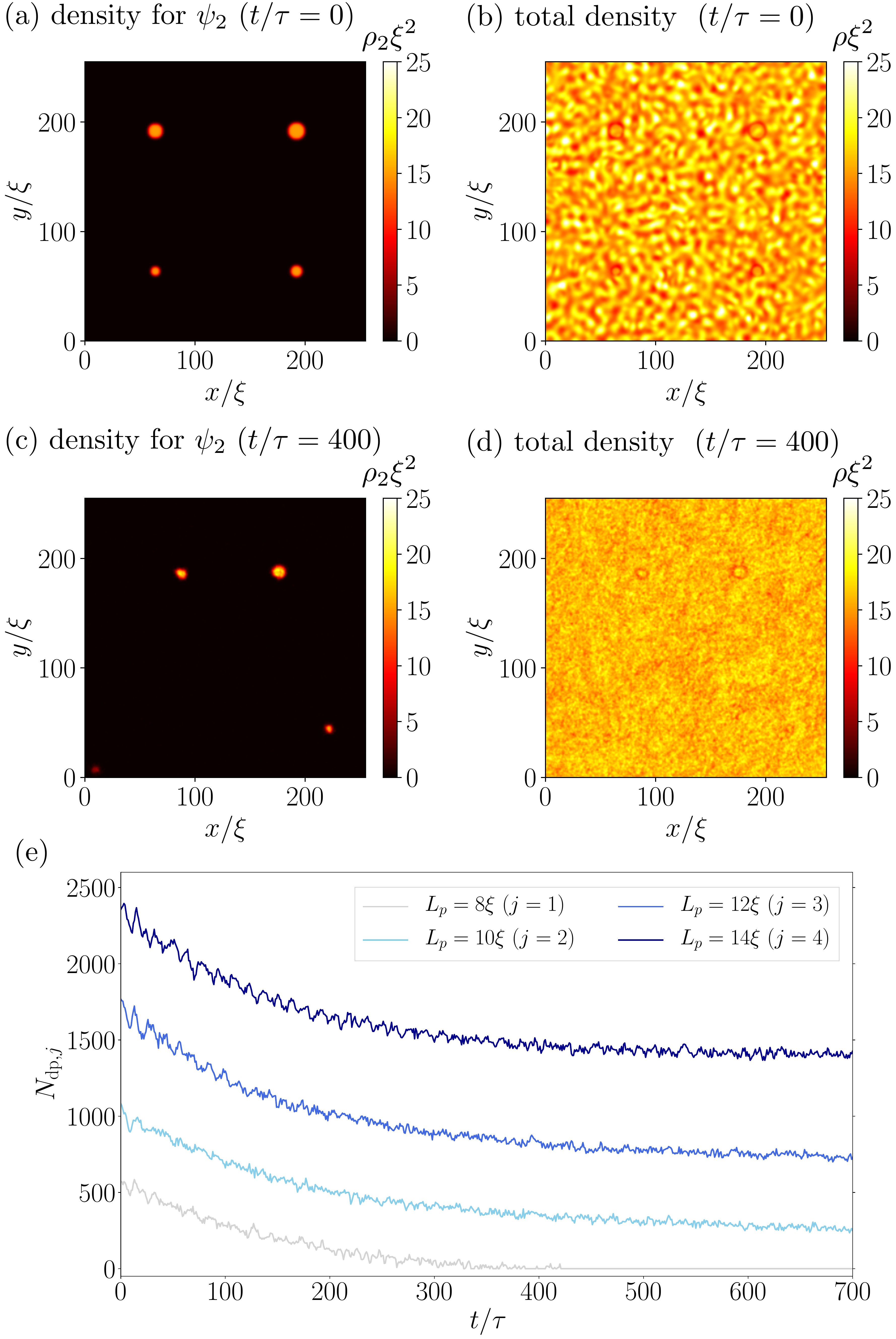}
\caption{
(Color online) Dynamics of four isolated droplets with the sizes $L_{p}/\xi=8,10,12$, and $14$ without any noises in the component $2$.
Initial distributions for $\rho_{2}(\bm{r},t)$ and $\rho(\bm{r},t)$ are shown in panels (a) and (b). As opposed to the case of Fig.~\ref{fig6}, we add noises only to the component $1$ in the initial wavefunction. The time-evolved distributions at $t/\tau=400$ are shown in panels (c) and (d), and we find that the sizes of the droplets decrease. 
This is also confirmed in panel (e) showing the droplet particle numbers $N_{{\rm dp},j}~{j=1, \ldots ,4}$.  
\label{fig7} }
\end{center}
\end{figure}

\subsubsection{Droplet stability mechanism}
We now give an essential insight for the droplet stability, and argue that fluctuations of the component $2$ can be emitted from or absorbed into droplets, and that the in- and out-flow balance supported by the two processes makes droplets stable. 

To investigate this droplet stability mechanism, we perform a numerical simulation starting from the following initial state without any noise terms in the component $2$:
\begin{eqnarray}
\psi_1(\bm{r},0) &=& \psi_{1}^{\rm (dp)}(\bm{r}) \nonumber \\
&+& \frac{1}{\sqrt{S}} \sum_{\bm{k} \in \{ \bm{k} | E_{\bm{k}} \leq \mu, \bm{k} \neq 0 \} } \bigl( u_{\bm{k}}\beta_{1,\bm{k}} + v_{\bm{k}}\beta_{1,-\bm{k}}^{*}  \bigl) e^{i\bm{k}\cdot\bm{r}},  \nonumber \\
\label{noise3_1}
\end{eqnarray}
\begin{eqnarray}
\psi_2(\bm{r},0) = \psi_{2}^{\rm (dp)}(\bm{r}). \label{noise3}
\end{eqnarray}
Figures~\ref{fig7}(a)--(d) shows time evolution of $\rho_2(\bm{r},t)$ and $\rho(\bm{r},t)$, and we find that the droplets become smaller in time. 
Correspondingly, the domain particle numbers within $\mathcal{B}_{j}$ exhibit clear decreases as in Fig.~\ref{fig7}(e).
Thus, the domains show diffusive behavior by emitting their particles.

\begin{figure}[t]
\begin{center}
\includegraphics[keepaspectratio, width=8.6cm,clip]{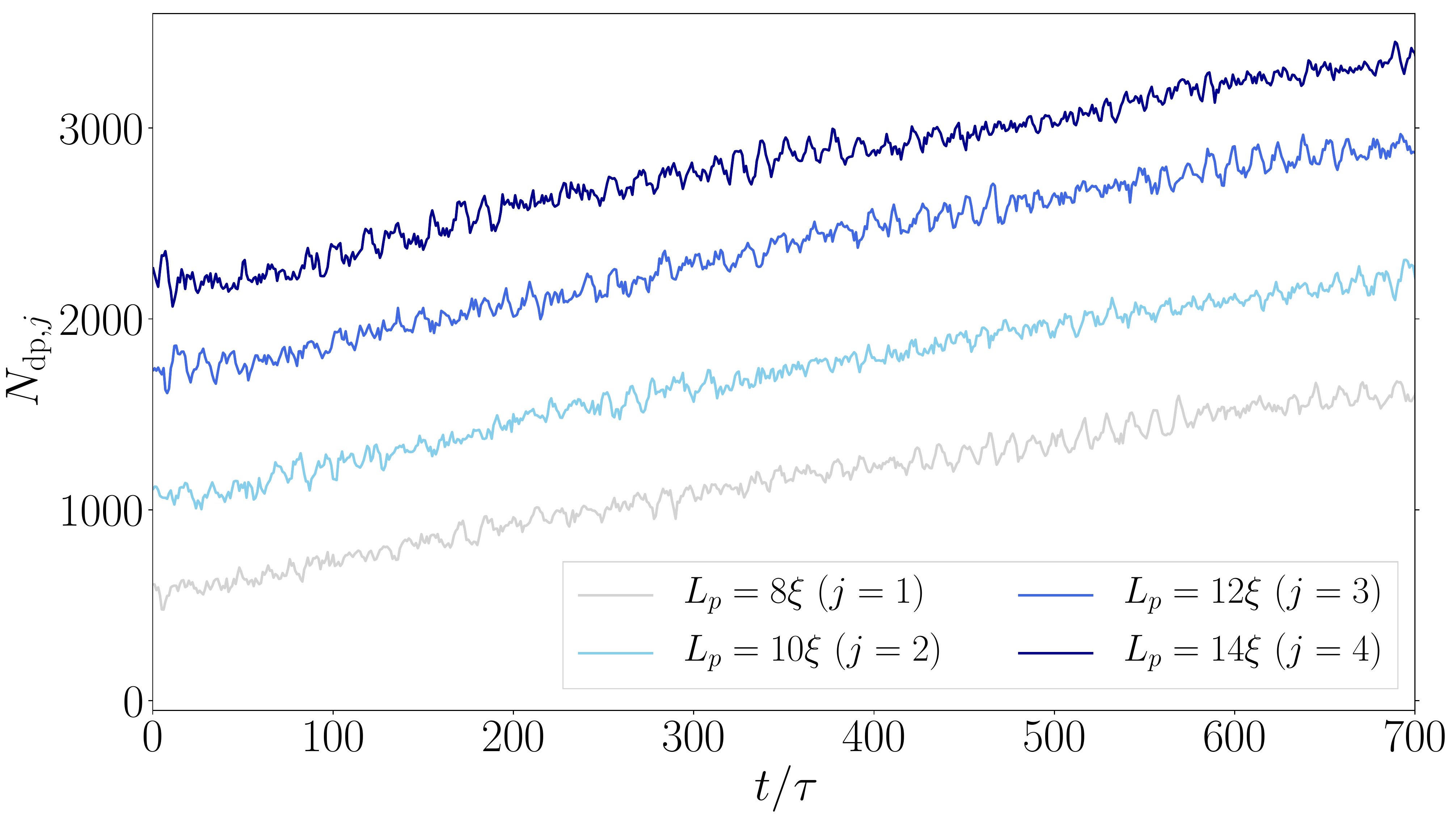}
\caption{
(Color online) Domain particle numbers under the strong noises in the component $2$. The numerical parameters used here are same as those in Fig.~\ref{fig6} except for $b_2=1.5$. All the droplets grows in time due to the large fluctuations of $\rho_2(\bm{r},t)$ around them. 
\label{fig8} }
\end{center}
\end{figure}

Considering the above numerical results with the two different initial states [Eqs.~\eqref{noise2_1}, \eqref{noise2},  \eqref{noise3_1}, and \eqref{noise3}], we expect that the droplets in Fig.~\ref{fig6} are stabilized by balance between emission and absorption of the fluctuations of $\rho_2(\bm{r},t)$. This description is consistent with our results so far. In the simulation of Fig.~\ref{fig7}, the component-$2$ particles outside the droplets do not exit, and thus the absorption process is almost absent before fluctuations emitted from other droplets spread over space. As a result, the droplets continue to emit their particles by the fluctuations of the component $1$, and the sizes decrease in time. On the other hand, in the simulation of Fig.~\ref{fig6}, there exist the component-$2$ particles around the droplets as noises, both the emission and absorption processes can work well, and thus the isolated droplets become stable due to the balance. We also perform a numerical simulation with large noises in the component $2$, where we use $b_2=1.5$ larger than $0.6$ in Eq.~\eqref{noise2} and other parameters are same as those in Fig.~\ref{fig6}. If our argument is correct, the fluctuations of the component $2$ are excessively absorbed into the droplets, and their sizes should grow over time. As shown in Fig.~\ref{fig8}, we in fact confirm the droplet growth, which supports the droplet stability mechanism mentioned above. 

In classical dissipative systems, fluctuations around droplets vanish due to a diffusion process, and this kind of stability mechanism is not valid. However, because our isolated system conserves the total energy, the fluctuations survive for a long time and stabilize the droplets. 

Finally, we comment on possible scenarios of droplet collapse. If strength of the fluctuations is strong enough, droplets can collapse. However, in our quench dynamics, this process may be rare because this occurs only when a droplet size is comparable to the coherence length $\xi$. In fact, as shown in Fig.~\ref{fig6}, the droplet with $L_p=8\xi$, which is comparable to the smallest domain size in the quench dynamics, is still stable. As another mechanism, quantum fluctuations may break droplets. Several 1D spin system actually finds the domain melting \cite{domian_melt}. However, compared to the 1D system, quantum fluctuations should not be dominant in our case because the system considered in this work is a weakly interacting BEC. Thus, we do not expect that such domain melting often occurs in 2D binary BECs, although our numerical method with the truncated Wigner approximation-like noises does not fully include the effect of quantum fluctuations.

\subsection{ $1/3$ power law in conventional classical systems and the relation with our results}
Relaxation dynamics considered here has been investigated for a long time in terms of Ostwald ripening and off-critical quench in binary mixtures \cite{bray2002,onuki2002,Ostwald0,Ostwald1,Ostwald2,Ostwald3,Ostwald4,Ostwald5,Ostwald6}, in which the correlation length has been found to obey the $1/3$ power law. The exponent is identical to our analytical result \eqref{kinetic5} in the early time. In this section, we argue that the fundamental mechanism behind the $1/3$ power in the binary BECs is essentially different from that in classical systems. 

Let us briefly review a physical origin of the $1/3$ in classical systems. One standard explanation for Ostwald ripening is given by the Lifshitz-Slyozov-Wagner (LSW) theory \cite{Ostwald0,Ostwald1}, which predicted that small droplets shrink and vanishes while large ones grow in time because of diffusion processes of concentration fields around the droplets. Based on this evaporation and concentration, the LSW theory analytically derives the $1/3$ power law, and many numerical simulations for several models have confirmed it \cite{Ostwald2,Ostwald3,Ostwald4,Ostwald5,Ostwald6}.  

However, the classical mechanism of the $1/3$ power law does not emerge in binary BECs because of the energy conversation in Eqs.~\eqref{GP1} and \eqref{GP2}. Although in the LSW theory diffusion processes of fields play a pivotal role in relaxation of classical mixtures, the conservation law basically prohibits such process in the atomic BEC. Our model, actually, does not have diffusive terms breaking the energy conservation. Due to the conservation law, a droplet itself is quite stable and cannot shrink in our relaxation dynamics. In fact, this behavior is numerically confirmed in Sec.~\ref{SD}. As an exceptional process, too small droplets comparable to the coherence length $\xi$ can collapse by emitting the partilces when the background fluctuations are quite strong. However, this kind of collapse is rare as mentioned in Sec.~\ref{SD}. Thus, we conclude that the $1/3$ power in the binary BECs is fundamentally different from the conventional LSW mechanism.

We comment on the previous works \cite{Random1,Random2,Random3,Random4,Random5,Random6}, where random motion of droplets in classical systems is considered and various power exponents $\beta$ are analytically obtained. This analytical approach is similar to our calculation in Sec.~\ref{ana_derivation}, but most of the models used in previous literature have diffusive terms breaking the energy conservation and furthermore the droplets are not driven by the spatial gradient of spin and velocity. Thus, the fundamental relaxation mechanisms seem to be different from that in our systems. It is interesting to consider whether or not such droplet dynamics in classical systems still emerges in ultracold quantum gases, and it remains as a future work.

\subsection{Dependence of the dynamics on $R$}
We discuss how the relaxation dynamics changes depending on the particle-number imbalance parameter $R$. 
Our analytical model in Sec.~\ref{ana_derivation} is based on the stable droplet structures and their random collisions. 
It is obvious that such description should break down when $R$ is close to unity since large scale domains have surface excitations with various length-scales.
However, our numerical simulations with $R \sim 1$ tend to show the power law behavior similar to Eq.~\eqref{kinetic5}.
In this current work, we cannot sufficiently understand the behavior, and we leave it in the future work. 

\begin{figure}[t]
\begin{center}
\includegraphics[keepaspectratio, width=8.6cm,clip]{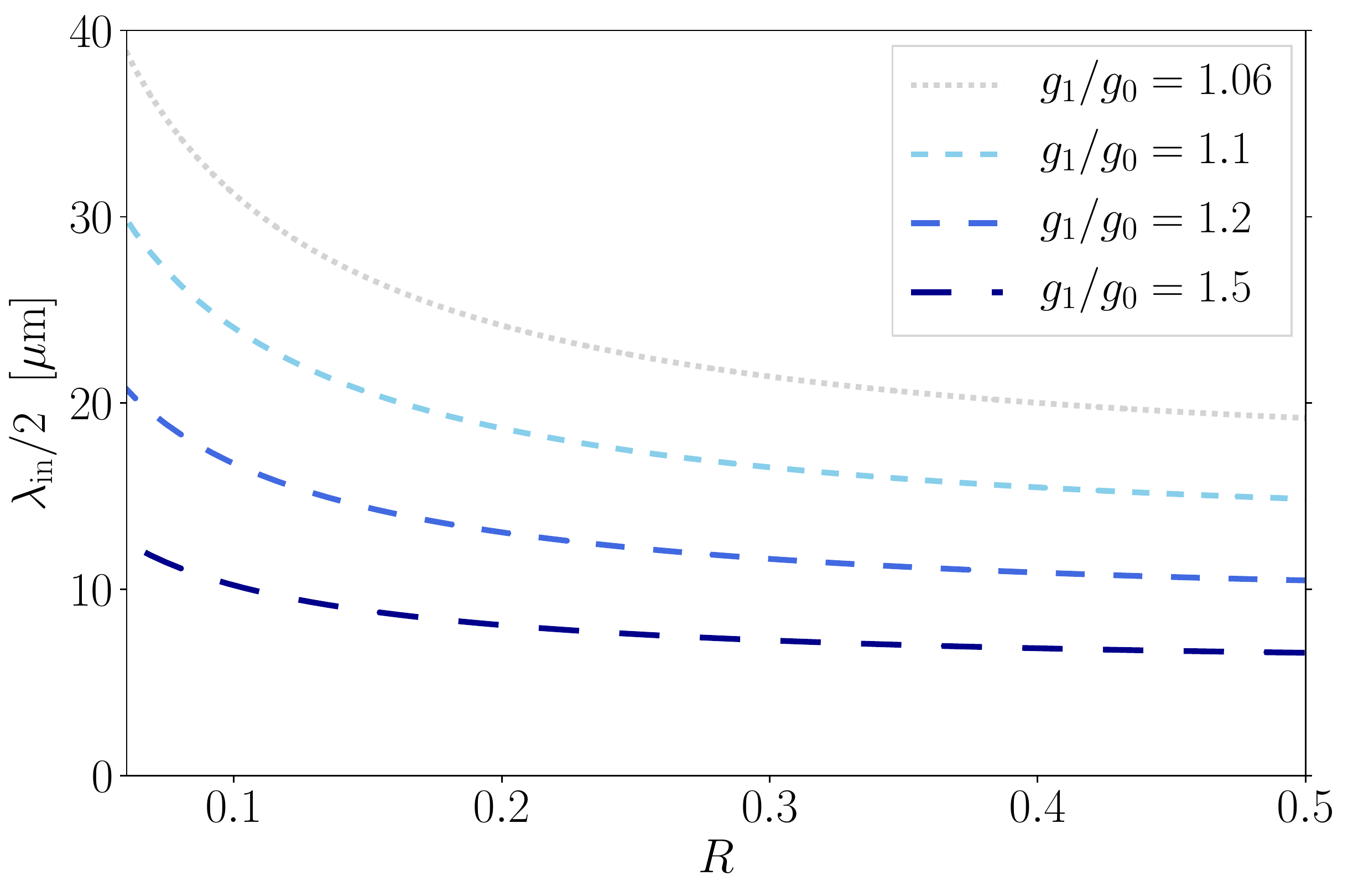}
\caption{(Color online) Dependence of the unstable wavelength \eqref{wavelength} on $g_1/g_0$ and $R$. The parameter used in the calculation is $\xi=0.5 {\rm \mu m}$.
\label{fig9} }
\end{center}
\end{figure}

\subsection{Possible experimental setups}
We discuss experimental possibility for observing our theoretical predictions. As described in the previous section, the universal scale-invariant dynamics is induced by random collisions between tiny droplet structures, so that we need to prepare a situation in which a droplet size is much smaller than a system size just after the tuning protocol. A diameter of a system in several experiments \cite{Shin1,Shin2} is about $200 ~{\rm \mu m}$ at the most, so that an averaged droplet linear-size is required to be smaller than at least $20~{\rm \mu m}$. The typical size is roughly estimated as $\lambda_{\rm in}/2$ in Eq.~\eqref{wavelength}. We plot its dependence on $g_1/g_0$ and $R$ in Fig.~\ref{fig9} using the typical coherence length $\xi=0.5~{\rm \mu m}$. We finds that the size can be smaller than $20~{\rm \mu m}$ if the ratio $g_1/g_0$ of the interaction couplings is larger than $1.1$. 

We consider two internal states $|F=1,m=-1 \rangle$ and $|  F=2,m=-2\rangle$ of $^{87}{\rm Rb}$ as a candidate to realize the suitable system mentioned above because the binary condensate has a long lifetime. This kind of system has in fact been used to study long-time dynamics \cite{Engels1}. The interaction couplings, however, are almost the same and the ground state is a miscible state. When we can tune the interaction coupling between the different components by means of the Feshbach resonance, our theoretical prediction should be observed.

\section{Conclusion \label{Conclusion}}
We have theoretically studied the relaxation dynamics in binary BECs with large particle-number imbalances, and uncovered the emergence of the scale-invariant relaxation dynamics. The significance of our finding is that the relaxation dynamics exhibits the dynamical scaling of the correlation function and the scaling law have two different power exponents depending on the elapsed time. We have analytically derived the $1/3$ and $1/2$ power laws for the growth of the correlation length, predicting the crossover in time and then numerically confirm it. As described in the derivation, the origin of the power laws is random collisions between droplets, whose centroids fluctuate with size-dependent velocities. The size dependence is determined by which is the main contribution to the  spin current, the spatial variation of the spin density vector or the mass current. The former (latter) is dominant for a small (large) droplet size and leads to $1/3~(1/2)$ power law. Hence, the crossover from the $1/3$ to $1/2$ power law occurs as the droplet size grows. Our prediction can be experimentally realized by utilizing the spin polarized mixture of $F=1$ and $2$ hyperfine states of $^{87}{\rm Rb}$ atoms with a controlled inter-species interaction by means of the Feshbach resonance.

\begin{acknowledgments}
This work was supported by JST-CREST (Grant No. JPMJCR16F2) and JSPS KAKENHI Grant Nos. JP15K17726, JP18K03538, JP19H01824, and JP19K14628.
\end{acknowledgments}

\appendix

\setcounter{equation}{0}
\setcounter{figure}{0}
\setcounter{section}{0}
\renewcommand{\theequation}{A-\arabic{equation}}
\renewcommand{\thefigure}{A-\arabic{figure}}

\section{Imaginary part of the correlation function \label{imaginary}}
As mentioned in Sec.~\ref{correlation_review}, the imaginary part of $C(r,t) $ is zero. Here, we give the proof in the following. 
The wavefunction can be expressed in the wave number space:
\begin{eqnarray}
\psi_2({\bm x},t) =  \sum_{\bm{k}} \bar{\psi}_2({\bm k},t) e^{i \bm{k} \cdot \bm{x}}
\label{imaginary2}
\end{eqnarray}
with the Fourier component $\bar{\psi}_2({\bm k},t)$. Substituting Eq.~\eqref{imaginary2} into Eq.~\eqref{correlation1}, we obtain 
\begin{eqnarray}
C(r,t) &=& \sum_{\bm{k}} |\bar{\psi}_2({\bm k},t)|^2 \bigl{ \langle} e^{-i \bm{k} \cdot \bm{r}} \bigl{\rangle}_{\Omega} \nonumber \\
&=& \sum_{\bm{k}} |\bar{\psi}_2({\bm k},t)|^2 J_0(kr)
\label{imaginary3}
\end{eqnarray}
with the Bessel function $J_n(x)~(n=0,1,2,\cdots$) of the first kind. Here, we calculate the angle-averaged part as
\begin{eqnarray}
 \bigl{ \langle} e^{-i \bm{k} \cdot \bm{r}} \bigl{\rangle}_{\Omega} &=& \frac{1}{2\pi} \int_{0}^{2 \pi} e^{-ikr {\rm cos \theta}} d \theta \nonumber \\
 &=& J_0(kr).
\label{imaginary4}
\end{eqnarray}
Thus, from Eq.~\eqref{imaginary3}, we find that the imaginary part of $C(r,t)$ becomes zero. 

\setcounter{equation}{0}
\setcounter{figure}{0}
\renewcommand{\theequation}{B-\arabic{equation}}
\renewcommand{\thefigure}{B-\arabic{figure}}
 
\section{Initial instability in the Bogoliubov theory \label{instability}}
We derive the expression of Eq.~\eqref{wavelength} for the most unstable wavelength and describe the strength of the instability. 
Applying the conventional Bogoliubov theory to Eqs.~\eqref{GP1} and \eqref{GP2} with $\Omega =0$ and a fixed particle-number imbalance $R$, we derive the Bogoliubov excitation energy \cite{pethick}:
\begin{eqnarray}
E(k)^2 = \epsilon(k) ( \epsilon(k) - A ), 
\label{instability1}
\end{eqnarray}
where
\begin{eqnarray}
\epsilon(k) = \frac{\hbar^2k^2}{2M}, 
\label{instability2}
\end{eqnarray}
\begin{eqnarray}
A = - \mu + \sqrt{  g_0^2 (\rho_1 - \rho_2)^2 + 4 g_1^2 \rho_1 \rho_2  }
\label{instability3}
\end{eqnarray}
with $\rho_1 = \rho_{\rm tot}/(1+R) $ and $\rho_2 = \rho_{\rm tot}R/(1+R)$. 
The wavenumber $k_{\rm in}$ determined by $\epsilon(k_{\rm in}) = A/2$ gives the largest imaginary part of $E(k)$ due to the quadratic form with respect to $\epsilon(k)$ in Eq.~\eqref{instability1} when the phase-separation condition $g_0 < g_1$ is satisfied. The analytical expression for $k_{\rm in}$ becomes
\begin{eqnarray}
k_{\rm in} = \frac{\sqrt{MA}}{\hbar}, 
\label{instability4}
\end{eqnarray}
which leads to Eq.~\eqref{wavelength} through $\lambda _{\rm in } = 2 \pi / k_{\rm in}$.

Next, we give an expression for the most largest imaginary part of $E(k)$. Substituting Eq.~\eqref{instability4} into Eq.~\eqref{instability1}, we obtain
\begin{eqnarray}
|{{\rm Im} E(k_{\rm in})}| &=& \frac{A}{2} \nonumber \\
&=& \frac{1}{2} \Biggl( - \mu + \frac{\mu}{1+R} \sqrt{  (1-R)^2 + \frac{4R g_1^2 }{g_0^2}  } \Biggl). \nonumber \\
\label{instability5}
\end{eqnarray}
When the parameters $R$, $g_0$, and $g_1$ satisfy $g_0 < g_1$ and $0 < R < 1$, the amplitude $|{{\rm Im} E(k_{\rm in})}|$ is a monotonically increasing function of $R$. This readily confirmed by
\begin{eqnarray}
&&\frac{d}{dR}|{{\rm Im} E(k_{\rm in})}| = \frac{ \displaystyle  \mu (1-R) \biggl( \frac{g_1^2}{g_0^2} - 1 \biggl) }{ (1+R)^2 \sqrt{ \displaystyle (1-R)^2 + \frac{4R g_1^2}{g_0^2}  }} > 0.
\nonumber  \\
\label{instability6}
\end{eqnarray}
Thus, we find that the initial instability becomes stronger when $R$ is larger.
As described in the caption of Fig.~\ref{fig5}(a), our numerical simulations are consistent with this dependence.

\setcounter{equation}{0}
\setcounter{figure}{0}
\renewcommand{\theequation}{C-\arabic{equation}}
\renewcommand{\thefigure}{C-\arabic{figure}}

\section{Derivation of the continuity equation for $f_z$ \label{fz_con}}
This appendix derives Eqs.~\eqref{spin1} and \eqref{spin2} from the GP equations \eqref{GP1} and \eqref{GP2} with $\Omega = 0$. 
The time derivative of $f_z$ in the GP model readily leads to 
\begin{eqnarray}
\frac{\partial}{\partial t} \rho f_{z} + {\bm \nabla} \cdot {\bm J}_{z} = 0, \label{AA1}
\end{eqnarray} 
\begin{eqnarray}
{\bm J}_{z} =  \frac{\hbar}{2Mi} \sum_{m,n=1}^{2} (\sigma_z)_{mn} \bigl( \psi_{m}^* \bm{\nabla} \psi_{n} - \psi_{n} \bm{\nabla} \psi_{m}^*  \bigl). \label{AA2}
\end{eqnarray} 

To rewrite Eq.~\eqref{AA2} in terms of the spin density vector $f_{\nu}$, we use the following notation for the macroscopic wavefunctions:
\begin{eqnarray}
\begin{pmatrix}
\psi_1 \\
\psi_2  
\end{pmatrix}
= \sqrt{\rho} e^{i \Phi}
\begin{pmatrix}
{\rm cos } (\theta /2) e^{-i\phi/2} \\
{\rm sin } (\theta /2) e^{i\phi/2}, 
\end{pmatrix}
\label{AA3}
\end{eqnarray} 
where $\Phi$ is the overall phase and $\theta$ and $\phi$ are the polar and azimuthal angles of the spin density vector, i.e., $\bm f$ is given by
\begin{eqnarray}
&&f_x = {\rm sin}\theta ~{\rm cos}\phi, \\
&&f_y = {\rm sin}\theta ~{\rm sin}\phi, \\
&&f_z = {\rm cos}\theta.
\label{AA5}
\end{eqnarray} 
Substituting Eq.~\eqref{AA3} into Eqs.~\eqref{AA2} and \eqref{spin4}, we obtaion
\begin{eqnarray}
&&\bm{J}_z = \frac{\hbar \rho}{M} \biggl(  {\rm cos}\theta~\bm{\nabla} \Phi - \frac{1}{2} \bm{\nabla} \phi \biggl), \label{AA4_1} \\ 
&&\bm{v} = \frac{\hbar}{M} \biggl(   \bm{\nabla} \Phi - \frac{1}{2} {\rm cos} \theta ~\bm{\nabla} \phi    \biggl). 
\label{AA4_2}
\end{eqnarray} 
By calculating the spatial gradient of $f_x$ and $f_y$, the following equality is obtained:
\begin{eqnarray}
f_x \bm{\nabla} f_y - f_y \bm{\nabla} f_x = {\rm sin}^2 \theta~\bm{\nabla} \phi.
\label{AA6}
\end{eqnarray} 
We substitute Eqs.~\eqref{AA4_2} and \eqref{AA6} into Eq.~\eqref{AA4_1}, and then obtain
\begin{eqnarray}
{\bm J}_{z} = \rho \biggl[ f_z \bm{v} + \frac{\hbar}{2M}  \Big( f_y {\bm \nabla} f_x - f_x {\bm \nabla} f_y  \Big)\biggl]. \label{AA7} 
\end{eqnarray}

\setcounter{equation}{0}
\setcounter{figure}{0}
\renewcommand{\theequation}{D-\arabic{equation}}
\renewcommand{\thefigure}{D-\arabic{figure}}

\section{Relation between $L_{\rm dp}(t)$ and $L_{\rm md}(t)$ \label{proportion}}
In Sec.~\ref{grwoth_law}, we have introduced two length scales $L_{\rm md}(t)$ and $L_{\rm dp}(t)$ as the mean-droplet distance and the typical linear size of droplets, respectively. In this Appendix, we show that they are related to each other due to the particle number conservation. Suppose that we have $N_{\rm dp}(t)$ droplets with the typical linear size $L_{\rm dp}(t)$. Then, the imbalance parameter $R$ is estimated as
\begin{eqnarray}
R= \frac{N_2}{N_1} = \frac{ \eta L_{\rm dp}(t)^2 N_{\rm dp}(t)}{S -  \eta L_{\rm dp}(t)^2 N_{\rm dp}(t)}
\end{eqnarray}
with a constant $ \eta \sim \mathcal{O}(1)$. 
The mean distance $L_{\rm md}(t)$ between the droplets can be expressed by
\begin{eqnarray}
L_{\rm md}(t) =  \sqrt{ \zeta \frac{ S - \eta L_{\rm dp}(t)^2 N_{\rm dp}(t) }{ N_{\rm dp}(t) } }
\end{eqnarray}
with a constant $\zeta \sim \mathcal{O}(1)$.
Then, combining the above equations, we obtain
\begin{eqnarray}
L_{\rm dp}(t) = \sqrt{ \frac{R}{\eta \zeta} } L_{\rm md}(t).
\end{eqnarray}

\bibliography{reference}

\end{document}